\documentclass[10pt]{article}
\usepackage{cite}
\usepackage{geometry}
\usepackage{graphicx}
\usepackage{slashed}
\usepackage{amsmath}
\usepackage{verbatim}
\usepackage{url}
\usepackage{amsfonts}
\usepackage{layout}
\usepackage{slashed}
\usepackage{dcolumn}
\usepackage{bm}

\usepackage{amssymb}
\usepackage{psfrag}
\def\tQ{\tilde{Q}}

\newcommand{\myperp}{{{\perp}}}
\newcommand{\bMHV}{\overline{\text{MHV}}}
\newcommand{\sep}{\makebox[.15cm]{$]\hskip-.070cm[$}}

\newcommand{\massA}{\rotatebox{12.5}{\boldmath$\delta$\hskip-.035cm}}

\newcommand{\barW}{{\overline{W\!}\,}}

\newcommand{\A}{{\cal A}}

\renewcommand{\baselinestretch}{1.05}
\newcommand{\dash}{\text{-}}

\newcommand{\eps}{\epsilon}

\def\d{\delta}

\newcommand{\ie}{{\it i.e.}\ }

\newcommand{\cn}{{\cal N}}
\newcommand{\be}{\begin{equation}}
\newcommand{\bea}{\begin{eqnarray}}
\newcommand{\beq}{\begin{equation}}

\newcommand{\ee}{\end{equation}}
\newcommand{\eea}{\end{eqnarray}}
\newcommand{\eeq}{\end{equation}}
\newcommand{\lsim}{\!\mathrel{\hbox{\rlap{\lower.55ex \hbox{$\sim$}} \kern-.34em \raise.4ex \hbox{$<$}}}}
\newcommand{\gsim}{\!\mathrel{\hbox{\rlap{\lower.55ex \hbox{$\sim$}} \kern-.34em \raise.4ex \hbox{$>$}}}}

\newcommand{\reef}[1]{(\ref{#1})}
\newcommand{\<}{\langle}
\renewcommand{\>}{\rangle}

\begin{document}

\begin{titlepage}

\begin{flushright}
RUNHETC-2011-08\\
MCTP-11-17 \\
PUPT-2370 \\
\end{flushright}
\vspace{3.0cm}

\begin{center}
{\Large \bf Massive amplitudes on the Coulomb branch of $\cn=4$ SYM}

\vspace{0.2in}
{\bf Nathaniel Craig$^{a,b}$, Henriette Elvang$^{c}$, Michael Kiermaier$^d$ and Tracy Slatyer$^a$}\\
\vspace{0.2cm}
{\it $^a$Institute for Advanced Study\\ Princeton, NJ 08540}\\
\vspace{0.2cm}
{\it $^b$Rutgers University\\ Piscataway, NJ 08854}\\
\vspace{0.2cm}
{\it $^c$Michigan Center for Theoretical Physics (MCTP)\\Department of Physics, University of Michigan, \\ Ann Arbor, MI 48109}\\
\vspace{0.2cm}
{\it $^d$Department of Physics, Princeton University \\ Princeton, NJ 08544}\\
\end{center}

\vspace{0.4cm}

\begin{abstract}

We initiate a systematic study of amplitudes with massive external particles on the Coulomb-branch of $\cn=4$ super Yang Mills theory:
1) We propose that (multi-)soft-scalar limits of massless amplitudes at the origin of moduli space can be used to determine Coulomb-branch amplitudes to leading order in the mass.
 This is demonstrated in
 numerous
 examples.
2)  We find compact explicit expressions for several towers of tree-level amplitudes,
including scattering of two massive $W$-bosons with any number of positive helicity  gluons, valid for all values of the mass.
3) We present the general structure of superamplitudes on the Coulomb branch. For example, the
 $n$-point
``MHV-band'' superamplitude is proportional to a Grassmann polynomial of mixed degree 4 to 12, which is uniquely determined by supersymmetry. We find explicit tree-level superamplitudes for this MHV band and for
other simple sectors of the theory.
4) Dual conformal generators are constructed, and we explore the dual conformal properties of the simplest massive amplitudes.

Our compact expressions
for amplitudes and superamplitudes
should be 
of both theoretical and phenomenological interest; in particular the tree-level results carry over to truncations of the theory with less supersymmetry.
\end{abstract}

\end{titlepage}

\tableofcontents

\newpage

\setcounter{equation}{0}
\section{Introduction}

Planar on-shell scattering amplitudes of massless particles in $\cn=4$ SYM enjoy numerous remarkable properties: they are much simpler than Feynman rules indicate; they are well-behaved under both ordinary and dual superconformal symmetry~\cite{Drummond:2008vq,Brandhuber:2008pf}; they can be packaged  into superamplitudes that make these symmetries manifest~\cite{Drummond:2008bq,Drummond:2008cr,Brandhuber:2009xz,Elvang:2009ya,ArkaniHamed:2009vw};
and
  compellingly simple loop-order expressions have been obtained
both at the level of the integrand~\cite{ArkaniHamed:2010kv,Mason:2010yk} and of
the final integrated result~\cite{Goncharov:2010jf,Gaiotto:2011dt}.
Their good looks and  good behavior are likely due to the  underlying integrable structure of the planar sector of $\cn=4$ SYM. So is there any hope that scattering processes
 involving massive particles 
might enjoy  similar properties? --- or even  be simple? Introduction of massive particles breaks the conformal symmetry, and may well wreck the simplicity of amplitudes.
The goal of this paper is to show that many attractive results can  be achieved for tree-level amplitudes with massive particles, and that they arise from a natural connection to the massless amplitudes.

An ideal laboratory for studying amplitudes with massive external states is
 $\cn=4$ SYM on the Coulomb branch. To date, essentially all developments regarding on-shell scattering amplitudes in  $\cn=4$ SYM have focused on the theory at the origin of moduli space where all particles are massless. Recent work \cite{Alday:2009zm, Henn:2010bk,Henn:2010ir,Henn:2011xk} used states on the Coulomb branch to regularize IR divergences of the loop amplitudes, and in doing so it was shown that a version of dual conformal symmetry survives as long as the masses also transform appropriately. Until now, however, there has been no systematic study of Coulomb-branch amplitudes with massive external states.\footnote{That is not to say, of course, that no progress has been made; for preliminary discussions of scattering amplitudes on the Coulomb branch, see in particular \cite{Schabinger:2008ah, Boels:2010mj}.
 For the study of massive amplitudes using recursion relations in other theories, see for example~\cite{Dixon:2004za, Badger:2004ty, Berger:2006sh, Badger:2007si, Dixon:2009uk, Badger:2009hw,Boels:2008du,Boels:2009bv,Boels:2010mj,Badger:2005zh,Schwinn:2007ee}. 
  }
 In this work we  initiate such  a systematic analysis of massive tree-level amplitudes.

We push the theory onto the Coulomb branch by letting
 some of 
the scalars in the theory acquire a vacuum expectation value (vev); our choice of
vevs
leads to a higgsing of the
$U(N+M)$
gauge group to $U(N)\!\times\!U(M)$  and breaks the global  R-symmetry $SU(4)\to Sp(4)$.\,\footnote{
Most of our results immediately carry over to the more general symmetry-breaking pattern $U(\sum_i\! N_i)\!\to\! \prod_i\! U(N_i)$ for the gauge group, and $SU(4)_R\!\to\! SU(2)\!\times\!SU(2)$ for the R-symmetry.
}
 The massive $\cn=4$ multiplets contain W-bosons and their SUSY partners. Thus the familiar helicity amplitudes must be generalized to include massive external lines, and there will be new classes of amplitudes that vanish in the $m \to 0$ limit.
This includes
``ultra-helicity-violating'' (UHV) amplitudes with only one negative-helicity particle, as well as $SU(4)_R$-violating amplitudes.
Just as MHV amplitudes in massless $\cn=4$ SYM take an intriguingly simple explicit form, we derive simple all-order expressions for UHV and  maximally $SU(4)_R$-violating amplitudes using BCFW recursion~\cite{Britto:2004ap,Britto:2005fq}.

As an example of such massive amplitudes, let us present our result for the  all-$n$ tower of tree-level $W$-$\overline W$-gluon amplitudes:\footnote{
 All-$n$ results for certain amplitudes with two massive states have been given previously in the literature \cite{Forde:2005ue,Rodrigo:2005eu}, but in a different representation that involves quite elaborate sums. We thank M. Peskin for bringing this work to our attention.}
\bea
  \label{WWex}
 \bigl\< W^-_1 \barW^+_2\, g_3^+\, g_4^+ \cdots g_n^+\bigr\>
   &=&-\frac{m^2\<q\,1^\myperp\>^2\,[3|\prod_{i=4}^{n-1}(m^2-x_{i2}x_{2,i+1})|n]}{\<q\,2^\myperp\>^2\,\<34\>\<45\>\cdots \<n\!-\!1,n\>\prod_{i=4}^{n}(x_{2 i}^2+m^2)}\,.
\eea
Here we have introduced $x_{ij} = p_i + p_{i+1} +  \dots + p_{j-1}$. The $^{\perp}$ on the spinors of the massive lines 1 and 2 refers to the decomposition of their momenta
$p_i$ into two null directions, namely $p_i^\perp$ and a reference vector $q$.
This massive spinor helicity formalism was developed in \cite{Dittmaier:1998nn,Cohen:2010mi}, and we review its essentials in section \ref{s:towers}.
Note that the amplitude \reef{WWex} has familiar little-group scaling properties.

The concise form of the all-$n$ tower  \reef{WWex} suggests that the simplicity of amplitudes at the origin of moduli space persists as we venture onto the Coulomb branch. Indeed, there are several reasons why we expect amplitudes on the Coulomb branch of  $\cn=4$ SYM to be simple. One is, of course, the maximal supersymmetry, but there is another --- perhaps more interesting --- reason.
Since the masses are proportional to the scalar vevs, the small-mass\footnote{Throughout this paper, we consider masses to be small when they are small compared to the momentum-invariants of the scattering process.} limit takes us close to the theory  at the origin of moduli space; thus at least in this limit the massive amplitudes
 should
be simple.
 Indeed, the soft-momentum limits of massless scalars at the origin of moduli space probe the physics on the Coulomb branch. Hence
we propose
a
 \emph{precise 
connection between the  (multi-)soft-scalar  limits of the massless amplitudes at the origin of moduli-space and the Coulomb-branch amplitudes in the small-mass limit!}

We demonstrate this connection in several explicit examples. To illustrate the idea, consider the $n$-point amplitude
$\bigl\<W^- \,\barW^+\,\phi^{34}\,g^+\cdots g^+\bigr>_n$
of two conjugate massive $W$-bosons, one massless scalar and $n\!-\!3$ gluons. This amplitude vanishes in the massless limit where it is forbidden by $SU(4)$ R-symmetry; the broken R-symmetry
 allows it to be non-vanishing on the Coulomb branch.
For small mass, the leading $O(m)$-term can be reproduced exactly from the soft scalar limit $\eps \to 0$ of the massless $(n\!+\!1)$-point amplitude
$\bigl\<g^-\, \phi^{12}_{\eps q}\,\, g^+\,\phi^{34}\,g^+\cdots g^+\bigr\>_{n+1}$.
Furthermore, in the $\eps \to 0$ limit, the  $(n\!+\!1)$-point amplitude leaves behind information about the direction $q$ of the momentum of the scalar:  this $q$ is
 precisely
the reference vector introduced on the Coulomb branch
 to define a basis of polarization vectors for W bosons.
Thus we recover a nice physical interpretation
of
 the null vector $q$ 
that was originally introduced as a purely technical tool.
Our proposal is that $n$-point amplitudes with leading small-mass behavior  $O(m^s)$  match the symmetrized $s$-soft scalar limit of $(n\!+\!s)$-point amplitudes at the origin of moduli space.
This proposal is borne out by a variety of explicit examples.

It is natural 
to package
the massive amplitudes together
 using the unbroken
 $\cn=4$
supersymmetry of the Coulomb branch. Thus we commence the study of massive Coulomb-branch superamplitudes.\footnote{
 Since we want to recover the simplicity of massless amplitudes in the massless limit, we
choose
 a {\em chiral} representation of the superamplitude. This is to be contrasted with the non-chiral representations~\cite{Boels:2010mj,YutinHuang} that arise directly from a compactification of the 6d superamplitudes of~\cite{Bern:2010qa,Brandhuber:2010mm,Dennen:2010dh}.}
 The $W$-$\barW$-gluon amplitude \reef{WWex} vanishes in the massless limit, since the SUSY Ward identities for massless amplitudes
 forbid
vector amplitudes $\< ++ \dots +\>$ and $\< -+ \dots +\>$. For massive amplitudes, $\< ++ \dots +\>$
must still vanish in a helicity basis with only one reference vector $q$ \cite{Schwinn:2006ca}, but $\< -+ \dots +\>$ is allowed.
The ``ultra-helicity-violating'' (UHV) amplitudes
$\< -+ \dots +\>$
are therefore the simplest ones on the Coulomb branch, just like the Parke-Taylor amplitudes~\cite{Parke:1986gb} are the simplest ones at the origin of moduli space. Unlike amplitudes at the origin of moduli space, however, their encapsulation in superamplitudes is somewhat subtle. Recall that for the massless case, supersymmetry does not mix amplitudes in different N$^k$MHV sectors, so the SUSY and R-symmetry constraints can be solved
 independently
sector-by-sector, and the result can be encoded in superamplitudes of Grassmann degree $4(k+2)$. In contrast, the UHV amplitudes on the Coulomb branch correspond to superamplitude Grassmann polynomials of degree 4,
 which
does not itself close under supersymmetry.
 Instead it 
requires additional contributions of degree 6, 8, 10,  and 12. (We are assuming that an $SU(2)\times SU(2)$ subgroup of the $SU(4)$ R-symmetry is preserved, hence we admit only even orders in the Grassmann variables.) We call the resulting superamplitude the ``MHV-band'' since it
 reduces to the  familiar  MHV Grassmann delta-function  $\delta^{(8)}(\tilde{Q})$ in the massless limit.

We show that the combination of SUSY and $SU(2)\times SU(2)$ constraints determines the MHV band superamplitude completely up to an overall factor, which can be fixed by projecting out any amplitude; thus for the case of just two
 adjacent
massive states, we fix the entire MHV-band using the UHV amplitude \reef{WWex}. The result can be written compactly as
\bea
  \nonumber
    \A_n^{\rm MHV-band}
    &=&
    -
    \frac{[3|\prod_{i=4}^{n-1}[m^2\!-\!x_{i2}x_{2,i+1}]|n]}{\<1^\perp2^\perp\>^2\<34\>\<45\>\cdots \<n\!-\!1,n\>\prod_{i=4}^{n}(x_{2 i}^2\!+\!m^2)}
    \\[.5ex]
&&\times\biggl[\delta^{(4)}\!\bigl( |i^\perp\>\eta_{ia}\bigr)\!+\!\frac{m\<1^\perp2^\perp\>}{\<q1^\perp\>\<q2^\perp\>}\,\delta^{(2)}\!\bigl( \<qi^\perp\>\eta_{ia}\bigr)\biggr]^2\!\times
  \biggl[1-\frac{[1^\perp q][2^\perp q]}{m[1^\perp 2^\perp]}\,\delta^{(2)}\!
  \biggl( \frac{m_i\eta_{ia}}{[i^\perp q]}\biggr)
  \biggr]^2~~~~~~
  \label{MHVbandIntro}
\eea
for a particular choice\footnote{Here $q$ is constrained to satisfy $q\cdot(p_1+p_2)=0$. This choice leads to particularly simple expressions,
see
section~\ref{s:superA}.} 
 of the reference vector $q$.
The squares are understood as a product of two factors corresponding to the two $SU(2)$'s. This
 superamplitude includes
 amplitudes with any two particles from the massive multiplet on lines 1 and 2, 
 including longitudinal polarizations of the $W$-bosons, and we specify how to extract them. The superamplitude and the massive spinor helicity formalism can be encoded in  Mathematica, rendering it easy to project out any desired amplitude.

\begin{figure}
\begin{center}
\includegraphics[height=8cm]{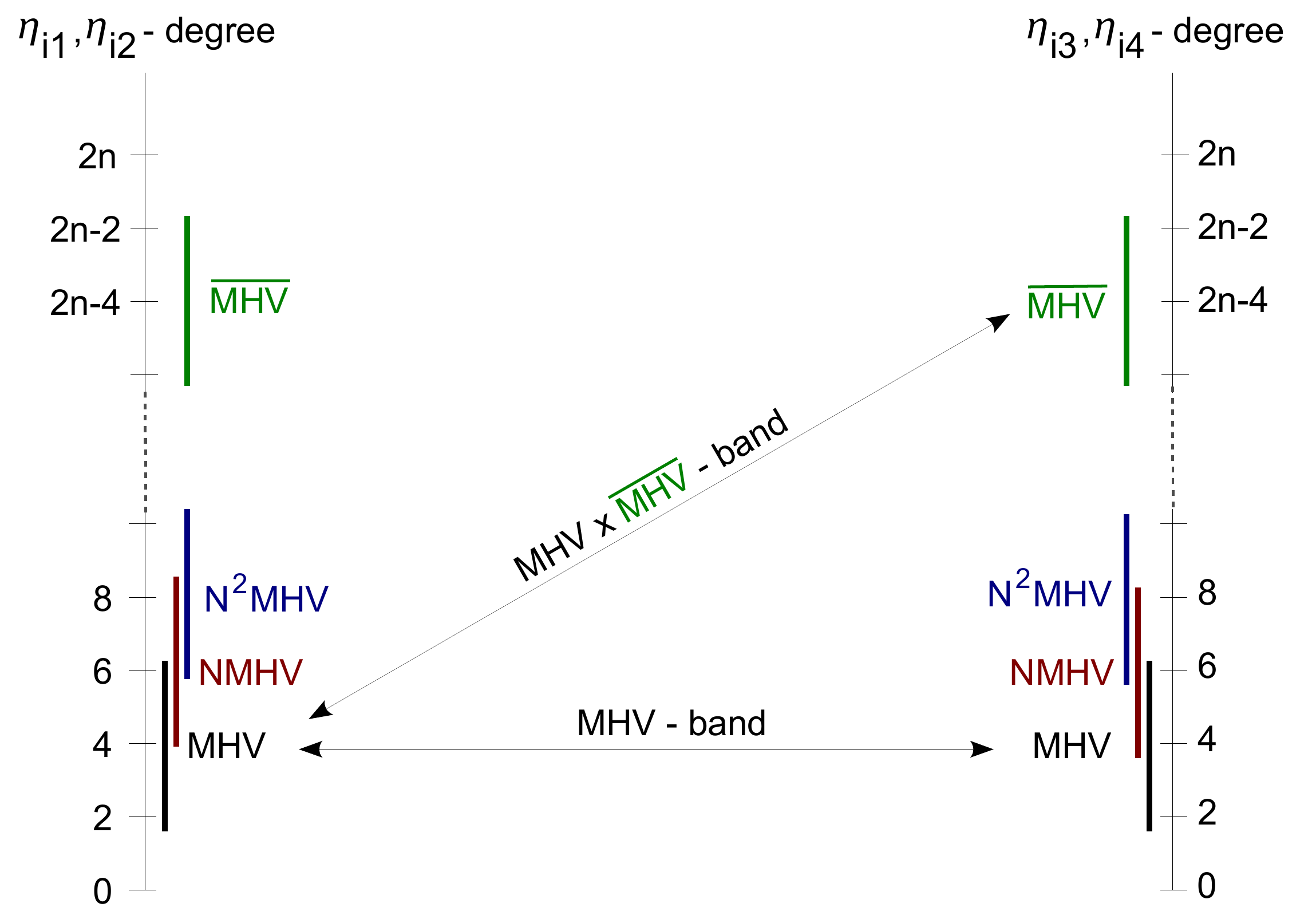}
\end{center}
\caption{The structure of superamplitudes on the Coulomb branch. We can assign a separate N$^k$MHV level to the two factors of $SU(2)$ corresponding to $\eta_{i1},\eta_{i2}$ and $\eta_{i3},\eta_{i4}$. The full Coulomb-branch superamplitude can then be decomposed into N$^k$MHV$\times$ N$^{k'}$MHV superamplitudes. }
\label{figsuperAmps}
\end{figure}

Given that the MHV-band superamplitude is an inhomogeneous $\eta$-polynomial of degree 4
 through 12, it is natural to wonder how the massive analogues of N$^k$MHV superamplitudes are structured. We find that these higher bands are inhomogeneous $\eta$-polynomials of increasing degree,
 but constant ``width''. Each band overlaps with its adjacent bands.
 For example, the NMHV-band involves $\eta$-polynomials of degree 8 through 16, so that it overlaps with
part of
 the MHV band.
 This means that
 this and  higher-bands must also be determined as needed in order to extract amplitudes associated with $\eta$-degrees 8 or higher. Also, the $SU(2)\times SU(2)$ R-symmetry structure allows us to assign distinct N$^k$MHV level to each of the two factors of $SU(2)$. At $6$-point, for example, there exist additional  MHV$\!\times\!$NMHV and MHV$\!\times\!\bMHV$ bands, which must vanish in the massless limit. The general superamplitude structure is illustrated in figure~\ref{figsuperAmps}.
We will discuss the SUSY structure in further detail
in section \ref{s:superA};
let us just mention here that the reason the sectors extend into bands is closely related to the fact that  the SUSY algebra on the Coulomb branch has a central charge \cite{Witten:1978mh,Fayet:1978ig}. The corresponding on-shell supercharges $Q^a$ and $\tilde{Q}_a$ annihilate the MHV-band superamplitude \reef{MHVbandIntro}.

Having established the band structure of superamplitudes on the Coulomb branch, we compute a variety of superamplitudes involving
 either two or
 arbitrary numbers of massive lines.
 In  particular, we present an extremely compact superamplitude expression for the maximally-$SU(4)_R$-violating MHV$\!\times\!\bMHV$ band with two adjacent massive lines, and fairly simple CSW-type expressions for all superamplitudes that are MHV or $\bMHV$ with respect to one of the two $SU(2)$ sectors. The latter is valid for any number of massive lines. We also present the special case of the $3$-point superamplitude; it turns out that supersymmetry alone fixes it completely --- including the relative normalization of the MHV, $\bMHV$ and MHV$\!\times\!\bMHV$ bands.
With these superamplitudes in hand, we
extend the connection between soft-scalar limits of massless amplitudes and small-mass limits of Coulomb-branch amplitudes to the full superamplitudes.

These excursions onto the Coulomb branch are all realized constructively in four dimensions. However, we may also recapture many of the same results in a somewhat different light by the reduction of maximal SYM in higher dimensions. In particular,
 $\cn=4$ SYM amplitudes on the Coulomb branch
 can be obtained
 through the dimensional reduction of $\cn = 1$ SYM in ten dimensions or $\cn = (1,1)$ SYM in six dimensions, both of which have been described in terms of a spinor helicity formalism \cite{Cheung:2009dc,CaronHuot:2010rj}. We focus on the 4d-6d connection, which suffices to capture all the Coulomb branch vacua preserving an $SU(2) \times SU(2)$ R-symmetry. Massless amplitudes in  $\cn=(1,1)$ SYM can be constructed using BFCW recursion relations, and at this time explicit results have been found for
 $3,4,5$-point
 superamplitudes~\cite{Bern:2010qa,Brandhuber:2010mm,Dennen:2010dh}.  
 The 4- and 5-components of the 6d  momenta are interpreted as (complex) masses in 4d: $p_4 \pm ip_5 = m_\pm$. Reduction of the 6d on-shell superfield formalism~\cite{Hatsuda:2008pm,Bern:2010qa,Brandhuber:2010mm,Dennen:2010dh}
 results in a non-chiral formulation of the 4d superamplitudes, but this
 can in principle be
 converted to the chiral formulation 
by a
 particular Grassmann 
Fourier transform~\cite{Hatsuda:2008pm,Elvang:2011fx}. 
We have verified numerically that the
4-point superamplitude in 6d matches the corresponding MHV-band superamplitude
 in 4d, and find in many cases that the 6d picture provides a natural organizing principle for our 4d results.

Finally, let us mention that  much progress on massless $\cn=4$ amplitudes was obtained from the Yangian structure, of which dual conformal symmetry constitutes one level. It is known that on-shell 
 tree
amplitudes of 6d $\cn=(1,1)$ have dual conformal symmetry \cite{Bern:2010qa,Brandhuber:2010mm,Dennen:2010dh} and that those of the Coulomb branch also do, again provided that the masses transform \cite{Alday:2009zm, Bern:2010qa}.
We probe the dual conformal transformation properties of amplitudes on the Coulomb branch and write some of our all-$n$ amplitudes
 in a manifestly dual conformal covariant way, albeit with ambiguous weights.

\medskip

{\em This paper is organized as follows.} In section~\ref{s:towers} we set up notation and present explicit examples for helicity amplitudes on the Coulomb branch. In particular, we present our results for UHV and maximally-$SU(4)_R$-violating amplitudes with adjacent massive legs for arbitrary $n$. In section~\ref{s:soft}, we present our proposal for the computation of Coulomb-branch amplitudes from the soft-scalar limits of massive amplitudes, which we illustrate with various examples. In section~\ref{s:superA}, we introduce Coulomb-branch superamplitudes; we discuss their general structure, how they simplify for a smart choice of reference vector $q$, and the special case of the 3-point superamplitude. We also present the more elaborate example of a match between the $s$-soft scalar limit of a
massless superamplitude with the leading order $m^s$ term of a massive amplitude, for any $s$. In section~\ref{sec:CSW}, we derive a CSW form~\cite{Cachazo:2004by} for the MHV$\!\times\!$N$^k$MHV band superamplitudes for arbitrary masses. In section~\ref{s:dci}, we discuss massive dual conformal
symmetry. 
In section~\ref{s:con} we
 outline avenues for future work.

\setcounter{equation}{0}
\section{Explicit amplitudes on the Coulomb branch}
\label{s:towers}

In this section, we begin our study of amplitudes on the Coulomb branch of $\cn=4$ SYM with a set of explicit examples. These examples are chosen to illustrate the physical aspects that
distinguish Coulomb-branch amplitudes from the amplitudes at the origin of moduli space, for instance R-symmetry breaking and the presence of longitudinal vector bosons.

To set the stage, consider the brane picture with a stack of $(N\!+\!M)$ D3-branes~\cite{Alday:2007hr,Alday:2009zm}. Let us separate $M$ branes from the others; obviously this breaks the $U(N\!+\!M)$ gauge group to $U(N) \!\times\! U(M)$. In this paper, we consider the simplest scenario with
scalar
vevs
\begin{equation}
\begin{split}
  \label{VEVs}
  \big\<(\phi^{12})_I{}^{J}\big\> &= \big\<(\phi^{34})_I{}^{J}\big\>  = v \, \delta_{I}{}^{J}
  ~~\text{ for }I,J\in U(M)\,,\\[1ex]
  \big\<\phi^{ab}\big\>~&=~0\hskip2.6cm\text{ otherwise}\,.
\end{split}
\end{equation}
The $\cn=4$ supersymmetry is preserved, but the global R-symmetry $SU(4)$ is broken to $Sp(4)$. The resulting spectrum has 5 Goldstone bosons.
The massive $\cn=4$ supermultiplets (arising from strings stretched between the separated branes) consist of bifundamentals of $U(N) \times U(M)$: massive W-bosons (3 d.o.f.), their wino partners $\psi$ (8 d.o.f.) and 5 massive scalars $w$. We can illustrate the splitting by writing the matrix fields in block-diagonal form
\bea
  \label{matrixform}
    (\hat A_\mu)=
    \bigg(
      \begin{array}{cc}
       (A_\mu)_{N\times N} &  (W_{\mu})_{N\times M}  \\[1mm]
         (\overline{W}_{\mu})_{M\times N}  & (\tilde{A}_\mu)_{M\times M} \\
      \end{array}
    \bigg)\, ,
\eea
and similarly for the other fields.
The $A_\mu$ and $\tilde{A}_\mu$ gluon multiplets
of $U(N)$ and $U(M)$
remain massless, while the
bifundamentals
$W$ and $\overline{W}$ are massive.
The masses are given by $m^2 = g^2 v^2$.
 In the following, we will suppress all dependence on the coupling $g$; in particular, we will set $\<\phi^{12}\>=v=m$.

The goal of this section is to present examples of explicit $n$-point amplitudes with massive states on the Coulomb branch.
Since the masses can be understood as momenta in the directions transverse to the branes, non-vanishing on-shell amplitudes must have $\sum_i m_i = 0$, where the sum runs over the external states.
In the
present
section, we focus on amplitudes with 2 adjacent external particles from the massive multiplet and $n-2$ from the familiar massless multiplet. The resulting trace-structure of the color-ordered amplitude is easily inferred from the block-matrix form.

To find compact results for the amplitudes, we adapt here the massive spinor-helicity formalism of~\cite{Dittmaier:1998nn},
 using the notation of~\cite{Cohen:2010mi}:
we introduce a light-like reference
vector $q$ and decompose the massive momenta $p_i$ as\footnote{
One can choose a different reference spinor $q_i$ for each line; however, the amplitudes are significantly simpler when all $q_i$'s are equal, $q_i=q$. We make this choice throughout the paper.}
\begin{equation}\label{decpi}
    p_i~=~p_i^\myperp-\frac{m_i^2}{2q\cdot p_i}q
   \,, ~~~~~~~~\text{with}~~~~~
   p_i^2 = - m_i^2 \,,~~~~~
   (p_i^\perp)^2 = q^2 = 0
    \,.
\end{equation}
We then express amplitudes in terms of the spinors $|i^\myperp\>$, $|i^\myperp]$ and $|q\>$, $|q]$ associated with the null vectors $p_i$ and $q$.\footnote{
The little-group ambiguity in the spinors of $p_i^\myperp$ results in $\<q i^\myperp\>=[i^\myperp q]$. This
condition
can simplify expressions for amplitudes, but the perp'ed spinors then no longer satisfy the conventional little-group transformation properties familiar from massless amplitudes. Throughout this paper, we will not use $\<q i^\myperp\>=[i^\myperp q]$ and instead keep conventional little-group properties of perp'ed spinors manifest.
See \cite{Cohen:2010mi} for further discussion of little-group properties of massive amplitudes.}
For massive vector bosons, it is convenient to use the following basis of polarization vectors:
\bea\label{pol}
  \nonumber
    \epsilon_-=\frac{\sqrt{2}|i^\myperp\> [q|}{[ i^\myperp q]}~~\,,~\qquad
    \epsilon_+=\frac{\sqrt{2}|q\> [i^\myperp|}{\< i^\myperp q\>}
    ~~\,,~\qquad
  \slashed{\epsilon}_0 &=&
   \frac{1}{m_i} \Big( \slashed{p}_i^\myperp  - \frac{m_i^2}{\<q|p_i|q]} \slashed{q} \Big)\,.
\eea
The amplitudes we present are ``helicity amplitudes'' in the basis of these polarizations.\footnote{For simplicity, we refer to the polarization $\epsilon^\pm$ of~(\ref{pol}) as positive/negative helicity both in the massless and massive case.}
 Helicity of the massive particles is only well-defined in a fixed Lorentz-frame;
 different choices of $q$ lead to physically distinct polarizations, and generically the helicity amplitudes
 depend on the reference spinor $q$. In  the massless limit, $q$-independence is recovered.

In this spinor-helicity formalism, 3-point
amplitudes with transverse vectors
take a simple form:
\begin{equation}\label{3point}
    \big\<W^-\overline{W\!}\,^+g^+\bigr\>=\frac{[2^\myperp3]^4}{[1^\myperp2^\myperp][2^\myperp3][3\,1^\myperp]}\,,\qquad~~
\big\<W^+\overline{W\!}\,^+g^-\bigr\>=\frac{[1^\myperp 2^\myperp]^4}{[1^\myperp2^\myperp][2^\myperp3][3\,1^\myperp]}\,,
\end{equation}
and similarly for their conjugates.
The
$W$ and $\barW$
vector bosons
have masses $m_W=-m_{\overline{W\!}}\equiv m$, and $g$ is a  massless gluon. We recognize  the conventional Parke-Taylor amplitudes, with regular spinors replaced by perp'ed spinors on the massive lines. The massless limit is easily recovered by removing the perp's.

We focus first on amplitudes with transverse (positive/negative) polarizations and discuss longitudinal vector bosons shortly thereafter. To further illustrate the structure of  amplitudes on the Coulomb-branch, we give the following $4$-point examples:
\begin{equation}
\begin{split}
~~
  \big\<W^-\overline{W\!}\,^-g^+g^+\bigr\>~=~-\frac{\<1^\myperp2^\myperp\>^2[34] }{\<34\>(P_{23}^2+m^2)} \,,
  ~~~~~~~
    \big\<W^-\overline{W\!}\,^+g^+g^+\bigr\>~=~-\frac{m^2\<q1^\myperp\>^2[34]}{\<q2^\myperp\>^2\<34\>(P_{23}^2+m^2)}\,.
    \qquad
\end{split}
\label{4ptWW}
\end{equation}
In the massless limit, the first amplitude
reduces to
the Parke-Taylor expression $\<12\>^3/\<23\>\<34\>\<41\>$. We note that the mass enters through the perp'ed spinors as well as in the propagator $1/(P_{23}^2 +m^2)$.

The second amplitude of \reef{4ptWW} is ``ultra-helicity violating'' (UHV) and vanishes at the origin of moduli space, $m=0$, as a consequence of the supersymmetry constraints for amplitudes with massless particles.
With massive external particles and a single reference vector $q$,
amplitudes with just positive helicity vectors
vanish identically \cite{Schwinn:2006ca}, 
as in the massless case.
However, the supersymmetric Ward identities in the massive case allow vector amplitudes with one negative-helicity particle to be non-vanishing \cite{Schwinn:2006ca}.
These ultra-helicity-violating (UHV) amplitudes, together with their supersymmetric cousins, comprise the simplest sector of amplitudes on the Coulomb branch; their simplicity is analogous to that of MHV amplitudes in the massless case.

UHV amplitudes can be systematically computed for any distribution of masses on the external legs; we demonstrate this in section \ref{s:superA}. Here we concentrate on the case of only two adjacent massive W-bosons $\<W^-\barW^+g^+\dots g^+\>$. The $n$-point formula is derived recursively from a gluonic $[3,4\>$ BCFW shift, with the 3-point amplitude $\<W^-\barW^+g^+\>$  from~(\ref{3point}) as input. We obtain the astonishingly {\em simple
all-$n$
form for the UHV-sector
tree
amplitude}
\begin{equation}
\boxed{\phantom{\Biggl(}
\begin{split}\label{WWalln}
\bigl\< W^-_1 \barW^+_2\, g_3^+\, g_4^+ \cdots g_n^+\bigr\>
   ~&=~-\frac{m^2\<q1^\myperp\>^2\,[3|\prod_{i=4}^{n-1}(m^2-x_{i2}x_{2,i+1})|n]}{\<q2^\myperp\>^2\,\<34\>\<45\>\cdots \<n\!-\!1,n\>\prod_{i=4}^{n}(x_{2 i}^2+m^2)}\,,
\end{split}
~}
\end{equation}
where we defined $x_{ij}\!=\!p_i\!+\!p_{i+1}\!+\!\dots\!+\!p_{j-1}$. It is interesting to consider this amplitude both in the small and large mass limit. To leading order in {\em small mass}, we have
\begin{equation}\label{WWallnsmallm}
  \bigl\< W^-_1 \barW^+_2\, g_3^+\, g_4^+ \cdots g_n^+\bigr\>~=~\frac{m^2\<q1^\myperp\>^2\<1^\myperp 2^\myperp\>}{\<q2^\myperp\>^2\,\<2^\myperp 3\>\<34\>\<45\>\cdots \<n\!-\!1,n\>\<n\,1^\perp\>}~+~O(m^4)\,.
\end{equation}
In section \ref{s:soft}, we show how this leading-order term can be obtained from double-soft scalar limits of amplitudes at the origin of moduli space.

In the {\em large-mass limit}, on the other hand, we obtain
\begin{equation}
\bigl\< W^-_1 \barW^+_2\, g_3^+\, g_4^+ \cdots g_n^+\bigr\>
   ~=~-\frac{\<q1^\myperp\>^2\,[3n]}{\<q2^\myperp\>^2\,\<34\>\<45\>\cdots \<n\!-\!1,n\>}+O(1/m^2)\,.
\end{equation}
In this limit, the amplitude can be interpreted as a solution to self-dual Yang-Mills theory in the background created by the heavy $W$-boson,
 which can be analyzed, for example, with the methods of 
\cite{Bardeen:1995gk,Selivanov:1996gw,Korepin:1996mm}.

In massless amplitudes, $SU(4)$ invariance dictates that each $SU(4)$ index $a=1,2,3,4$ must occur the same number of times on the external states of any non-vanishing amplitude. For example, the amplitude $\<\phi^{12}\phi^{12}\phi^{34}\phi^{34}\>$ is non-vanishing, while the amplitude $\<\phi^{12}\phi^{34}\phi^{34}\phi^{34}\>$ vanishes in the massless theory. With $SU(4)_R$ symmetry broken to $Sp(4)$ on the Coulomb branch, the pairs of $SU(4)$ indices $\{1,2\}$ and $\{3,4\}$ can appear in different multiplicities.
The maximal $SU(4)$ violation occurs in amplitudes that contain the indices $\{1,2\}$ only once, but contain  $n\!-\!1$ instances of the indices
\{3,4\}.
 At the
3-point
 level, an example of such an amplitude is
\begin{equation}\label{WWphi}
    \big\<W^-\,\overline{W\!}\,^+\phi^{34}\bigr\>
    ~=~ - m\, \frac{\<1^\perp|q|2^\perp]}{\<2^\perp|q|1^\perp]}\,.
\end{equation}
For general $n$, the sector of {\em maximally $SU(4)$-violating amplitudes}
includes the amplitude tower
\begin{equation}\label{WWphis}
    \bigl\< W^-_1 \barW^+_2\, \phi^{34}_3\, \phi^{34}_4 \cdots \phi^{34}_n\bigr\>
   ~=~-\frac{m^{n-2}\<1^\perp|q|2^\perp]}{\<2^\perp|q|1^\perp]\prod_{i=4}^{n}(x_{2 i}^2+m^2)}\,.
\end{equation}
This result  was derived recursively from a $[3,4\>$ shift, starting with $\<W^- \barW^+ \phi^{34}\>$ given in~(\ref{WWphi}), though this particular amplitude is so simple that it can also be directly computed from Feynman diagrams to all $n$.

Next consider
{\em longitudinal vector bosons}. On the Coulomb branch, the gluons `eat' a scalar to become the massive $W$-bosons. In terms of the familiar massless labeling of the scalars, the longitudinal mode of the $W$-boson can thus be identified as $W^L=(w^{12}\!+\!w^{34})/\sqrt{2}$. The orthogonal linear combination $w^\perp=(w^{12}\!-\!w^{34})/\sqrt{2}$ is one of the five scalars in the massive multiplet; the other four are $w^{13}$, $w^{14}$, $w^{23}$, and $w^{24}$. Thus if we write on-shell helicity amplitudes in terms of $w^{12}$ and $w^{34}$ (which is technically convenient) we must keep in mind that they  contain both longitudinal gauge-boson and scalar  components. For this reason, such amplitudes typically depend explicitly  on the reference spinor $q$.
As examples, let us present the following 3-point amplitudes:
\begin{equation}
\begin{split}
  \label{3pointB}
    \big\<W^L\,\overline{W\!}\,^-g^+\bigr\>\,&=\,\sqrt{2}\,\big\<w^{12}\,\overline{W\!}\,^-g^+\bigr\>=-\sqrt{2}m\frac{\<2^\myperp\!|q|3]^2}{\<q|1|q]\<q|2|q]}\,,\\
    \big\<W^{L}\overline{W\!}\,^{L}\phi^{34}\bigr\>&=\frac{1}{2}\Bigl(\big\<w^{12}\overline w^{12}\phi^{34}\bigr\>+\big\<w^{12}\overline w^{34}\phi^{34}\bigr\>+\big\<w^{34}\overline w^{12}\phi^{34}\bigr\>\Bigr) 
    =-\frac{m}{2}\,\frac{\<q|1|q]^2+\<q|2|q]^2+\<q|3|q]^2}{\<q|1|q]\<q|2|q]}\,.
\end{split}
\end{equation}
These amplitudes all vanish in the massless limit, because they violate the $SU(4)$ R-symmetry. They can be non-vanishing when $m\ne 0$ because they respect the unbroken $SU(2) \times SU(2) \subset Sp(4)$ R-symmetry.
The  amplitudes \reef{3pointB} are related to the amplitude~(\ref{WWphi}),
and in fact also to the $SU(4)$-preserving amplitudes~(\ref{3point}),
 by supersymmetry (see section \ref{s:3ptsuperA}).

%%%%%%%%%%%%%%%%
%%%%%%%%%%%%%%%%
%%%%%%%%%%%%%%%%
\setcounter{equation}{0}
\section{Constructing massive amplitudes from massless ones}
\label{s:soft}
In theories with spontaneously broken global symmetries, the flat directions of the vacuum manifold reveal themselves through the properties of amplitudes involving Goldstone bosons: the amplitudes vanish as the momentum of any Goldstone boson is taken soft. This property has many useful consequences, from its original discovery in the context of pion
physics
\cite{Adler:1964um}
to the study of finiteness in $\cn=8$ supergravity~\cite{Bianchi:2008pu,ArkaniHamed:2008gz,Kallosh:2008rr,Brodel:2009hu,Elvang:2010kc,Bossard:2010dq,Bossard:2010bd,Beisert:2010jx}. The scalars of $\cn=4$  SYM theory at the origin of moduli space, however, are of course not
Goldstone
bosons --- as we move out onto the Coulomb branch, the physics is genuinely different.
Thus we
do not expect the scalar soft limits of  amplitudes in massless SYM theory to vanish. Instead, their soft limits should allow us to probe the physics on the Coulomb branch. In this section we illustrate that the soft-scalar limits of massless amplitudes correctly reproduce the leading term in the small-mass expansion of Coulomb branch tree amplitudes.
It is trivial to reproduce the leading order of amplitudes that are $O(1)$ in the massless limit: to leading order,
these
massive and massless amplitudes simply coincide. We thus start by considering  Coulomb-branch amplitudes that are $O(m)$.

\subsection{$O(m)$ amplitudes from massless amplitudes}\label{3.1}
Consider the  Coulomb-branch amplitude $\<W^- \barW^+\phi^{34}g^+\cdots g^+\>$ of two massive $W$ bosons, a massless scalar, and arbitrarily many positive-helicity gluons.
This amplitude is $SU(4)$-violating and thus vanishes in the massless limit. To leading order in mass it can be derived using superamplitude techniques, as we show in section \ref{s:superA}. The result is
\begin{equation}
    \bigl\<W_1^- \barW_2^+\phi_3^{34}g_4^+\cdots g_n^+\bigr\>~=~
    -
    m\frac{\<q1^\perp\>
    \<1^\perp3\>^2
    }{\<q2^\perp\>\<2^\perp3\>\cdots\<n\,1^\perp\>}
    +O(m^3)\,.
    \label{WWphi34}
\end{equation}
We expect that this leading-order behavior of the amplitude is
accessible through scalar soft-limits of amplitudes at the origin of moduli space. But which massless amplitude should we pick? In particular, at what position(s)  should we insert a soft scalar into the amplitude? To answer this question, let us have a closer look at the trace structure of~(\ref{WWphi34}). Denoting $SU(N)$ indices by $A,B,\ldots$ and $SU(M)$ indices by $I,J,\ldots$, the trace structure takes the form
\begin{equation}
    (W^-)_A{}^I (\barW^+)_I{}^B(\phi^{34})_B{}^C\cdots (g^+)_D{}^A\,.
\end{equation}
The scalar we want to insert should be along the direction of the vev,
$\<(\phi^{ab})_I\!{\,}^J\>\!=\!m(\delta_{12}^{ab}+\delta_{34}^{ab})\delta_I\!{\,}^J$. There is only one place in the color trace where a scalar with color-index structure $\delta_I\!{\,}^J$ can fit: between the two  $W$ bosons. This leads us to consider the massless analogue of the amplitude~(\ref{WWphi34}), with an additional soft scalar along the vev direction inserted between the first two vectors. This massless $\cn=4$ amplitude is given by
\begin{equation}\label{singsoft}
    \bigl\<g_1^-\, \phi^{12}_{\epsilon q}\, g_2^+\,\phi_3^{34}\,g_4^+\cdots g_n^+\bigr\>_{n+1}~=~
    \frac{\<1 q\>^2\<1 3\>^2}{\<1 q\>\<q2\>\<23\>\cdots\<n1\>}\,.
\end{equation}
The corresponding amplitude with the other vev scalar $\phi^{34}$ inserted between the two gluons vanishes, because it violates $SU(4)$.
Comparing this to~(\ref{WWphi34}), we find
\begin{equation}\label{expropm}
  \bigl\<W_1^- \,\barW_2^+\,\,\phi_3^{34}\,g_4^+\cdots g_n^+\bigr>_n~=~
  m\lim_{\epsilon\to0}\bigl\<g_1^-\, \phi^{12}_{\epsilon q}\,\, g_2^+\,\phi_3^{34}\,g_4^+\cdots g_n^+\bigr\>_{n+1}
  +O(m^3)\,.
\end{equation}
It is intriguing that the soft-limit crucially depends on the {\em direction $q$}
along
which we take the scalar momentum soft.

Let us now formulate the lessons from the above example.
We claim that the leading term in $O(m)$ amplitudes can be obtained from the corresponding massless amplitude in which pairs of massive $W$ bosons $W_i$, $\barW_j$ are replaced by gluons $g_i$, $g_j$ and  soft scalars in the vev-direction are inserted between them.
The leading-order mass dependence comes from the vev(s)
(in the example above $\langle \phi_{12} \rangle = m$)
such that
 for 
 the case of 2 adjacent massive lines, we propose
\begin{equation}\label{proposalm}
    \boxed{\phantom{\biggl(}
    \bigl\<W_1\,\barW_2 \cdots\>_n~=~\<\phi_{ab}\>\,\lim_{\epsilon\to0}\,\bigl\<\,g_1\, \phi^{ab}_{\epsilon q}\,g_2\,\cdots\bigr\>_{n+1}+O(m^3)\,.
    }
\end{equation}
Similarly for other particles of the massive $W$-multiplets.
Some comments on this proposal are in order:
\begin{itemize}
  \item The right-hand side generically {\em depends on the direction $q$ along which we take the scalar momentum $p_\phi=\epsilon q$ to zero}. This $q$-dependence translates into the $q$-dependence of the polarization vectors~(\ref{pol}) on the left-hand side. Thus in this limit, the reference spinor $q$ in the massive spinor helicity formalism has a natural interpretation as the direction of the soft scalar momentum!

  \item As the gluons $g_1$ and $g_2$ on the right-hand side are massless, they cannot carry the massive momenta $p_1$ and $p_2$ of $W_1$ and $\barW_2$ on the left-hand side. Instead, we assign them the $q$-projected  momenta $p_1^\perp$ and $p_2^\perp$. Naively, this violates momentum conservation. However, momentum is still conserved to leading order in $m$, and the leading term on the right-hand side is thus unambiguous.\footnote{One could be pedantic and enforce momentum conservation by correlating the small-mass and $\epsilon\to0$ limit as $\epsilon=-\frac{m^2}{2q\cdot p_1}-\frac{m^2}{2q\cdot p_2}$. Then, $\sum p^\perp_i+\epsilon q=0$. However, the leading mass term is independent on how one takes the limit $\epsilon\to0$.}
    \item When the color-trace of the Coulomb-branch amplitude allows the insertion of a scalar along the vev directions in several positions, the right-hand side of~(\ref{proposalm}) turns into a sum over soft-scalar insertions.
    This can occur when there is more than one pair of $W$-bosons, and/or when a pair of $W$-bosons is not adjacent in the color trace.  We will illustrate this momentarily in an example.
\end{itemize}

\noindent
{\bf Example: Amplitudes with non-adjacent $W$-bosons}\\
Consider the amplitude $\<W^- \,\tilde{g}^+\,\barW^+\phi^{34}g^+\cdots g^+\>$,
where $\tilde{g}$ denotes a $U(M)$ gluon and $g$ are the usual $U(N)$ gluons.
To leading order, it is given by
\begin{equation}
    \bigl\<W_1^-\, \tilde{g}_2^+\,\barW_3^+\phi_4^{34}g_5^+\cdots g_n^+\bigr\>~=~
    -
    m\frac{\<q1^\perp\>\<1^\perp 3^\perp\>\<1^\perp4\>^2}{\<q3^\perp\>\<1^\perp 2\>\<23^\perp\>\<3^\perp4\>\cdots\<n\,1^\perp\>}
    +O(m^3)\,.
    \label{WgWphi34}
\end{equation}
The color-structure of this amplitude is
\begin{equation}
    (W^-)_A{}^I (\tilde{g}^+)_I{}^J (\barW^+)_J{}^B(\phi^{34})_B{}^C\cdots (g^+)_D{}^A\,.
    \label{nonadjW1}
\end{equation}
Since
the gluon between the two $W$'s carries indices in $U(M)$, there are now {\em two} places where we can insert the vev scalar:
between
 $W^-$ and $\tilde{g}^+$, or between $\tilde{g}^+$ and $\barW^+$. The corresponding contributions from the massless amplitudes can be computed as in~(\ref{singsoft})
\begin{equation}
\begin{split}
    \bigl\<g_1^-\, \phi^{12}_{\epsilon q}\,\,g_2^+\, g_3^+\,\,\phi_4^{34}\,g_5^+\cdots g_n^+\bigr\>_{n+1}~=~
    \frac{\<1q\>^2\<1 4\>^2}{\<12\>\<23\>\cdots\<n1\>}\times\frac{\<12\>}{\<1q\>\<q2\>}\,,\\
    \bigl\<g_1^-\, g_2^+\,\phi^{12}_{\epsilon q}\,\,g_3^+\,\,\phi_4^{34}\,g_5^+\cdots g_n^+\bigr\>_{n+1}~=~
    \frac{\<1q\>^2\<1 4\>^2}{\<12\>\<23\>\cdots\<n1\>}\times\frac{\<23\>}{\<2q\>\<q3\>}\,.
\end{split}
\end{equation}
Adding
the two contributions gives
\begin{equation} 
    \bigl\<g_1^-\, \phi^{12}_{\epsilon q} \, g_2^+ g_3^+\,\phi_4^{34}\,g_5^+\cdots g_n^+\bigr\>_{n+1}
    +\bigl\<g_1^-\, g_2^+\,\phi^{12}_{\epsilon q} \,g_3^+\,\phi_4^{34}\,g_5^+\cdots g_n^+\bigr\>_{n+1}
    ~=~-\frac{\<q1\>\<13\>\<1 4\>^2}{\<q3\>\<12\>\<23\>\cdots\<n1\>}\,.
\end{equation}
Comparing this to~(\ref{WgWphi34}) gives a precise match to our conjecture for this example.

We encountered in section \ref{s:towers}, eq.~\reef{WWalln}, a tower of amplitudes $\bigl\< W^-_1 \barW^+_2\, g_3^+\, g_4^+ \cdots g_n^+\bigr\>$ whose leading terms in small-mass were $O(m^2)$.
Now we match those
at leading order
by taking double-soft scalar limits from the origin of moduli space.

\subsection{$O(m^2)$ amplitudes from massless amplitudes}
To generalize our proposal to amplitudes whose leading term is $O(m^2)$,  two soft scalars must be inserted. Naively, one might propose
\begin{equation}\label{naive}
    \bigl\<W_1\,\barW_2 \cdots\>_n~\overset{?}=~\<\phi_{ab}\>\<\phi_{cd}\>\,\lim_{\epsilon\to0}\,\bigl\<g_1\, \phi^{ab}_{\epsilon q}\phi^{cd}_{\epsilon q}\,g_2\,\cdots\bigr\>_{n+2}+O(m^4)\,.
\end{equation}
This prescription indeed works
when the scalars do not have collinear divergences.
Consider, for example,
the small-mass limit of the
maximally $SU(4)$ violating 4-point amplitude
\begin{equation}
    \bigl\< W^-_1 \barW^+_2\, \phi^{34}_3\, \phi^{34}_4\bigr\>
   ~=~\frac{-m^{2}\<1^\perp|q|2^\perp]}{\<2^\perp|q|1^\perp](x_{2 4}^2+m^2)}~=~
   m^{2}\,\frac{-\<1^\perp|q|2^\perp]}{\<2^\perp|q|1^\perp]\<23\>[23]}+O(m^4)\,.
\end{equation}
For this example, the only non-vanishing massless amplitude that appears on the right-hand side of~(\ref{naive}) is the NMHV 6-point amplitude
$\< g^- \phi^{12}\phi^{12} g^+ \phi^{34} \phi^{34}\>$. After some algebra, which we present in more detail in section~\ref{s:multisoft}, we find
\begin{equation}
    \lim_{\epsilon\to 0}\bigl\< g^-_1 \,\phi^{12}_{\epsilon q}\,\phi^{12}_{\epsilon q} \,g^+_2\, \phi^{34}_3\, \phi^{34}_4\bigr\>~=~ -\frac{\<1|q|2]}{\<2|q|1]\<23\>[23]}\,, \label{6ptsoft}
\end{equation}
and thus
\begin{equation}
    \bigl\< W^-_1 \barW^+_2\, \phi^{34}_3\, \phi^{34}_4\bigr\>~=~m^2 \lim_{\epsilon\to 0}\bigl\< g^-_1 \,\phi^{12}_{\epsilon q}\,\phi^{12}_{\epsilon q} \,g^+_2\, \phi^{34}_3\, \phi^{34}_4\bigr\>+O(m^4)\,,
\end{equation}
in precise agreement with~(\ref{naive}).

Generically, however, the right-hand side of~(\ref{naive}) is {\em divergent} due to the collinear scalars. The collinear divergences  stem from Feynman diagrams in which $\phi^{ab}$ and $\phi^{cd}$ sit on the same 3-point vertex.\footnote{In the previous example, such 3-point interactions were prohibited by $SU(4)$-invariance.}
These divergences can be removed by simple symmetrization in the scalars $\phi^{ab}$ and $\phi^{cd}$. We thus propose, for $O(m^2)$ amplitudes,
\begin{equation}\label{proposalm2}
    \boxed{\phantom{\biggl(}
    \bigl\<W_1\,\barW_2 \cdots\>_n~=~\frac{1}{2}\<\phi_{ab}\>\<\phi_{cd}\>\,\lim_{\epsilon\to0}\lim_{q'\to q}\,
    \Bigl(\bigl\<g_1\, \phi^{ab}_{\epsilon q}\phi^{cd}_{\epsilon q'}\,g_2\,\cdots\bigr\>_{n+2}\,+\,
    \bigl\<g_1\, \phi^{cd}_{\epsilon q'}\phi^{ab}_{\epsilon q}\,g_2\,\cdots\bigr\>_{n+2}\Bigr)+O(m^4)\,.
    }
\end{equation}

As a concrete example, let us then consider the amplitudes $\<W^-\barW^+g^+\dots g^+\>$ in~(\ref{WWalln}). Its expression to leading order in mass was given in~(\ref{WWallnsmallm}). For this amplitude, the proposal~(\ref{proposalm2})
reads
\begin{equation}\label{propm2ex}
    \bigl\<W_1^-\barW_2^+g_3^+\cdots g_n^+\>_n~=~m^2\,\lim_{\epsilon\to0}\lim_{q'\to q}\,
    \Bigl(\bigl\<g_1^-\, \phi^{12}_{\epsilon q}\phi^{34}_{\epsilon q'}\,g_2^+\,g_3^+\cdots g_n^+\bigr\>_{n+2}\,+\,
    \bigl\<g_1^-\, \phi^{34}_{\epsilon q'}\phi^{12}_{\epsilon q}\,g_2^+\,g_3^+\cdots g_n^+\bigr\>_{n+2}\Bigr)+O(m^4)\,.
\end{equation}
 The right-hand side of~(\ref{propm2ex}) can be evaluated straight-forwardly:
\begin{equation}
\begin{split}\label{forsym}
&\lim_{\epsilon\to0}\lim_{q'\to q}\,
    \Bigl(\bigl\<g_1^-\, \phi^{12}_{\epsilon q}\phi^{34}_{\epsilon q'}\,g_2^+\,g_3^+\cdots g_n^+\bigr\>_{n+2}\,+\,
    \bigl\<g_1^-\, \phi^{34}_{\epsilon q'}\phi^{12}_{\epsilon q}\,g_2^+\,g_3^+\cdots g_n^+\bigr\>_{n+2}\Bigr)\\
    &=~\lim_{q'\to q}\frac{\<1 q\>^2\<1 q'\>^2}{\<23\>\<34\>\cdots\<n1\>}
    \biggl(\frac{1}{\<1 q\>\<qq'\>\<q'2\>}+\frac{1}{\<1 q'\>\<q'q\>\<q2\>}\biggr)\\
    &=~\frac{\<q1\>^2\<12\>}{\<q2\>^2\<23\>\<34\>\cdots\<n1\>}\,.
\end{split}
\end{equation}
Recalling the leading-mass expression (\ref{WWallnsmallm}) for $\<W^-\barW^+g^+\dots g^+\>$, this precisely confirms the claim~(\ref{propm2ex}).

Consider now an example with non-adjacent $W$-bosons. The
$n$-point Coulomb-branch amplitude  $\<W^-\tilde{g}^+\barW^+ g^+\cdots g^+\>$ is given by
\begin{equation}\label{nonadjmsquare}
    \bigl\<W^-\tilde{g}^+\barW^+ g^+\cdots g^+\bigr\>_n~=~m^2\frac{\<q1^\perp\>^2\<1^\perp3^\perp\>^2}{\<q3^\perp\>^2\<1^\perp 2\>\<2 3^\perp\>\cdots\<n 1^\perp\>}+O(m^4)\,.
\end{equation}
As in~(\ref{nonadjW1}),
there are again two places to insert vev-scalars: between $W^-$ and $\tilde{g}^+$, or between $\tilde{g}^+$ and $\barW^+$. As we have to distribute two scalars over these two places, there are now {\em three} distinct non-vanishing massless amplitudes to consider. When both scalars are inserted at the same location, we have
\begin{equation}
\begin{split}
\lim_{\epsilon\to0}\,
    \bigl\<g_1^-\, \phi^{12}_{\epsilon q}\,\phi^{34}_{\epsilon q}\,g_2^+\,g_3^+\cdots g_n^+\bigr\>_{\rm sym}
    ~&=~\frac{\<q1\>^2\<12\>}{\<q2\>^2\<23\>\cdots\<n1\>}\,,\\[.5ex]
\lim_{\epsilon\to0}\,
    \bigl\<g_1^-\, g_2^+\phi^{12}_{\epsilon q}\,\phi^{34}_{\epsilon q}\,g_3^+\cdots g_n^+\bigr\>_{\rm sym}
    ~&=~\frac{\<q1\>^4\<23\>}{\<q2\>^2\<q3\>^2\<12\>\<34\>\cdots\<n1\>}\,,
\end{split}
\end{equation}
where we denoted the symmetrization that was written out explicitly in~(\ref{forsym}) above by the subscript `sym', for simplicity.
Let us now insert the two scalars in distinct locations. The symmetrization is trivial in this case as the two terms are finite and identical. It gives
\begin{equation}
    \lim_{\epsilon\to0}\,\bigl\<g_1^-\, \phi^{12}_{\epsilon q}\,g_2^+\,\phi^{34}_{\epsilon q}\,\,g_3^+\cdots g_n^+\bigr\>_{\rm sym}~=~
    2\,\frac{\<q1\>^4}{\<1q\>\<q2\>\<2q\>\<q3\>\<34\>\cdots\<n1\>}\,.
\end{equation}
The sum of the three contributions allow us to complete a square and after a single   application of the Schouten identity, we obtain
\begin{equation}
\begin{split}
  \lim_{\epsilon\to0}\,&\biggl[
    \bigl\<g_1^-\, \phi^{12}_{\epsilon q}\,\phi^{34}_{\epsilon q}\,g_2^+\,g_3^+\cdots g_n^+\bigr\>_{\rm sym}
    +\bigl\<g_1^-\, g_2^+\phi^{12}_{\epsilon q}\,\phi^{34}_{\epsilon q}\,g_3^+\cdots g_n^+\bigr\>_{\rm sym}
    +\bigl\<g_1^-\, \phi^{12}_{\epsilon q}\,g_2^+\,\phi^{34}_{\epsilon q}\,\,g_3^+\cdots g_n^+\bigr\>_{\rm sym}
    \biggr]\\
    &=~\frac{\<q1\>^2\<13\>^2}{\<q3\>^2\<1 2\>\<2 3\>\cdots\<n 1\>}
    \,.
\end{split}
\end{equation}
Comparing this to~(\ref{nonadjmsquare}) again
demonstrates
a precise match.

%%%%%%%%%%%%%
%%%%%%%%%%%%%
\subsection{General $O(m^s)$ amplitudes}\label{s:genms}
For amplitudes whose leading term is $O(m^s)$,  it is now natural to conjecture
\begin{equation}\label{proposalmk}
    \bigl\<W_1\,\barW_2 \cdots\>_n~=~\frac{1}{s!}\<\phi_{a_1b_1}\>\cdots\<\phi_{a_sb_s}\>\,\lim_{\epsilon\to0}\,
    \bigl\<g_1\, \phi^{a_1b_1}_{\epsilon q}\cdots\phi^{a_s b_s}_{\epsilon q}\,g_2\,\cdots\bigr\>_{n+s, {\rm sym}}
    +O(m^{s+2})\,,
\end{equation}
where the subscript `sym' indicates the $s$-scalar generalization of the 2-scalar symmetrization given in~(\ref{proposalm2}).
We note that, for $s>2$, the only non-vanishing contributions to~(\ref{proposalmk}) arise from the case where the $\phi^{a_i b_i}$ are either all $\phi^{12}$ or all $\phi^{34}$.
If there were a non-vanishing amplitude containing mixed soft scalars such as $(\phi^{12})^{s-1}\phi^{34}$, then R-symmetry would also allow a non-vanishing soft-limit contribution with two fewer soft scalars, $(\phi^{12})^{s-2}$ --- but the latter would imply that $\bigl\<W_1\,\barW_2 \cdots\>_n$ is really $O(m^{s-2})$, not $O(m^{s})$, in contradiction with our attempt to extract the leading $O(m^s)$ term of the amplitude.\footnote{This argument is valid only for $s>2$. It does not apply to $O(m^2)$ dependence of UHV amplitudes derived above, which required a symmetrization of mixed scalars $\phi^{12}\phi^{34}$.  Indeed, R-symmetry does not forbid UHV amplitudes in the massless case, but they still vanish due to SUSY constraints. Therefore, UHV amplitudes on the Coulomb-branch are $O(m^2)$, not $O(1)$ as R-symmetry might have suggested.}
 With identical scalars, there are never any collinear divergences in~(\ref{proposalmk}), even before symmetrization. We conclude that for $s>2$,
\begin{equation}\label{proposalmkbetter}
    \bigl\<W_1\,\barW_2 \cdots\>_n~=~m^s\,\lim_{\epsilon\to0}\,\Bigl(\bigl\<g_1\, \phi^{12}_{\epsilon q}\cdots\phi^{12}_{\epsilon q}\,g_2\,\cdots\bigr\>_{n+s}+\bigl\<g_1\, \phi^{34}_{\epsilon q}\cdots\phi^{34}_{\epsilon q}\,g_2\,\cdots\bigr\>_{n+s}\Bigr)~+~O(m^{s+2})\,.
\end{equation} 
 In section~\ref{s:multisoft}, we present an explicit computation of the leading term in the $n$-point maximally-$SU(4)$-violating amplitude~(\ref{WWphis}) from soft limits of $n-2$ scalars. This computation,
carried out with the help of superamplitudes,
 gives a precise match
via
 the proposal~(\ref{proposalmkbetter}).

\setcounter{equation}{0}
\section{Coulomb-branch superamplitudes}\label{s:superA}
In this section we initiate the construction of superamplitudes for $\cn=4$ SYM on the Coulomb-branch.
We begin with the SUSY constraints, then study the form of the superamplitudes and their application to matching soft-scalar limits. The 4d-6d correspondence is briefly outlined at the end of the section.

%%%%%%%%%%%%%%%
In the massless case,
 the superalgebra of $\cn=4$ SYM is
\begin{equation}
    \bigl\{|\tQ_a\>^{\dot\alpha},[Q^b|^\beta\bigr\}~=~\delta_a^b\, p^{\dot\alpha\beta}\,,\qquad~~~
     \bigl\{[Q^a|^{\alpha},[Q^b|^\beta\bigr\}~=~\bigl\{|\tQ_a\>^{\dot\alpha},|\tQ_b\>^{\dot\beta}\bigr\}~=~0
     \,.
\end{equation}
It can be realized in the on-shell superfield formalism
by the following
operators
\begin{equation}
    |\tQ_a\>\,=\,|p\>\eta_{a}\,, \qquad~~~
         [Q^a|\,= \,[p|\frac{\partial}{\partial \eta_{a}}\,.
\end{equation}
On the Coulomb branch, we choose to give vevs
only to some of the scalars
\begin{equation}\label{central}
    \<\phi^{ab}\>~=~Z^{ab}~=~ Z^*_{ab}~=~  Z
  \begin{pmatrix}
     i\sigma_2 & 0 \\
    0 &  i\sigma_2
  \end{pmatrix}
  \,.
\end{equation}
Now the superalgebra  acquires a central charge:
\begin{equation}
    \bigl\{|\tQ_a\>^{\dot\alpha},[Q^b|^\beta\bigr\}~=~\delta_a^b\, p^{\dot\alpha\beta}\,,\qquad
     \bigl\{[Q^a|^{\alpha},[Q^b|^\beta\bigr\}~=~-\epsilon^{\alpha\beta}\, Z^{ab}\,,\qquad
    \bigl\{|\tQ_a\>^{\dot\alpha},|\tQ_b\>^{\dot\beta}\bigr\}~=~
    -\epsilon^{\dot\alpha\dot \beta}\,Z^*_{ab}\,.
\end{equation}
On superamplitudes, this algebra can be realized by the operators
\begin{equation}\label{Qs}
    [Q^a|\,= \,[p^\perp|\frac{\partial}{\partial \eta_{a}}-\frac{[q|}{[p^\perp q]}Z^{ab}\eta_{b}\,,
    \qquad~~ 
    |\tQ_a\>\,=\,|p^\perp\>\eta_{ia}-\frac{|q\>}{\<q p^\perp \>}Z_{ab}^*\frac{\partial}{\partial\eta_{b}}\,.
\end{equation}
The central charge $Z$ is related to the mass of $W$-bosons through $m^2=Z^*Z$. We will take $Z$ to be real in the following.

\subsection{MHV superamplitudes for $n>3$}
To write down Coulomb-branch superamplitudes, we 
need to determine the SUSY-invariant $\eta$-polynomial that generalizes the delta function $\delta^{(8)}(|i\>\eta_{ia})$ at the origin of moduli space.
This step is non-trivial because we are using the chiral representation of the superspace that is conventional at the origin of moduli space.
The supercharges~(\ref{Qs}) mix $\eta$ polynomials of different degrees; therefore, there is no polynomial of homogenous degree in $\eta$ that is annihiliated by these supercharges. The central charge~(\ref{central}) breaks $SU(4)\to Sp(4)
\supset
SU(2)\times SU(2)$, 
and
 it is convenient to determine SUSY invariants separately for the two $SU(2)$ sectors.
Explicitly, 
the supercharges for
the $\{1,2\}$ $SU(2)$-sector
take the form
\begin{equation}
\label{massive2comp12}
  \begin{split}
    &|\tQ_1\>\,=\,\sum_{i=1}^n\,|i^\perp\>\eta_{i1}\,-\,\bar\mu_i |q_i\>\frac{\partial}{\partial \eta_{i2}}\,,
    \qquad [Q^1|\,= \,\sum_{i=1}^n\,[i^\perp|\frac{\partial}{\partial \eta_{i1}}
    \,-\,\mu_i [q_i|\eta_{i2}
    \,,\qquad \\[1mm]
    &|\tQ_2\>\,=\,\sum_{i=1}^n\,|i^\perp\>\eta_{i2}\,+\,\bar\mu_i |q_i\>\frac{\partial}{\partial \eta_{i1}}\,,
    \qquad [Q^2|\,= \,\sum_{i=1}^n\,[i^\perp|\frac{\partial}{\partial \eta_{i2}}
    \,+\,\mu_i[q_i|\eta_{i1}
    \,,
      \end{split}
\end{equation}
where we defined
\begin{equation}
    \mu_i=\frac{m_i}{[i^\perp q]}\,,\qquad\bar\mu_i=\frac{m_i}{\<qi^\perp\>}\,.
\end{equation}

%%%%%%%%%%%%%%%

We now construct a SUSY invariant   annihilated by the SUSY charges $|\tQ_a\>$, $|\tQ^a\>$ with $a=1,2$.
We are looking for an invariant $\massA_{12}$ that generalizes the massless $\delta$-function,
so  it should  take the form
\begin{equation}
     \massA_{12}~=~\delta^{(4)}\!\bigl( |i^\perp\>\eta_{ia}\bigr)+O(m)\,.
\end{equation}
For any $n>3$,
this leads to
a unique SUSY invariant whose terms are of $\eta$-degree 6 or less.  It is given by
\begin{equation}
\begin{split}\label{massA12}
    \massA_{12}\,~=&\,~
  \delta^{(4)}\!\bigl( |i^\perp\>\eta_{ia}\bigr)\\
  &+  K_4\delta^{(2)}\!\bigl( \<qi^\perp\>\eta_{ia}\bigr)\delta^{(2)}\!\bigl( \mu_i\eta_{ia}\bigr)
+K_4'\Bigl[
  \delta^{(2)}\!\bigl( |i^\perp\>\eta_{i1}\bigr)\delta\bigl( \<qi^\perp\>\eta_{i2}\bigr)\delta\bigl( \mu_i\eta_{i2}\bigr)+
    \delta^{(2)}\!\bigl( |i^\perp\>\eta_{i2}\bigr)\delta\bigl( \<qi^\perp\>\eta_{i1}\bigr)\delta\bigl( \mu_i\eta_{i1}\bigr)
    \Bigr]
  \\[1ex]
  &
    +K_2\delta^{(2)}\!\bigl( \<qi^\perp\>\eta_{ia}\bigr)
      +K_6\delta^{(4)}\!\bigl( |i^\perp\>\eta_{ia}\bigr)\delta^{(2)}\!\bigl( \mu_i\eta_{ia}\bigr)\,.
\end{split}
\end{equation}
Here, $K_2$, $K_4$, $K_4'$ and $K_6$ are kinematic factors. For the case of two massive lines with masses $m_1=-m_2=m$, we have
\begin{equation}\label{2mKs}
\begin{split}
  K_2~=~\frac{mx_{13}^2}{\<q|1.2|q\>}\,,\qquad
  K_4~=~-\frac{[q|1.2|q]}{\<q|1.2|q\>}\,,\qquad
  K_4'~=~\frac{m\<q|x_{13}|q]}{\<q|1.2|q\>}\,,\qquad
  K_6~=~-\frac{\<q|1| q]\<q|2|q]}{m\<q|1.2|q\>}\,.
\end{split}
\end{equation}
Note that the factor of $1/m$ in $K_6$ is harmless because $K_6$ multiplies a term that is $O(\mu^2)\sim m^2$ in~(\ref{massA12}), which vanishes in the massless limit.
The general form of the kinematic factors $K_i$ with arbitrary number of
massive
lines 
will be given
in section \ref{supermultimassive}.

On the Coulomb-branch of $\cn=4$ SYM,  the conventional massless supermomentum delta-function generalizes to $\massA_{12}\times \massA_{34}$, where $\massA_{34}$ is simply $\massA_{12}$ with
R-symmetry
indices $a=1,2$ replaced by $a=3,4$. All superamplitudes (with $n>3$ legs) must contain a factor of $\massA_{12}\times \massA_{34}$. We will call the superamplitude in which the entire $\eta$-dependence is captured by $\massA_{12}\times \massA_{34}$ the ``MHV-band'' superamplitude; the MHV-band superamplitude contains terms with $\eta$-degrees ranging from 4 to 12. One amplitude in the sector of $\eta$-degree $4$ is sufficient to completely determine the MHV-band superamplitude. For the special case of two massive particles on lines 1 and 2, for example, the explicit tower of amplitudes~(\ref{WWalln})
determines the MHV-band superamplitude to be
\begin{equation}\label{AMHV}
    \boxed{
    \A_n^{\rm MHV-band}
    ~=~
    -
    \frac{[1^\perp2^\perp]^2[3|\prod_{i=4}^{n-1}[m^2-x_{i2}x_{2,i+1}]|n]}{\<34\>\<45\>\cdots \<n\!-\!1,n\>x_{13}^4\prod_{i=4}^{n}(x_{2 i}^2+m^2)}
    \times \massA_{12}\times \massA_{34}\,.}
\end{equation}

Let us now discuss the general structure of superamplitudes on the Coulomb-branch.
Beyond the MHV-band, it is convenient to first generalize our analysis within each $SU(2)$ sector. Invariants under the SUSY transformations~(\ref{massive2comp12}) can always be written as
\begin{equation}\label{NkMHVSU2}
    \A^\text{N$^k$MHV}~=~P_{12}^{k}\massA_{12}\,,
\end{equation}
where $P_{12}^{k}$ is a {\em homogeneous} polynomial of  degree $2k$ in $\eta_{i1}$, $\eta_{i2}$. The SUSY constraints on $P^k_{12}$ can be solved using the methods of~\cite{Elvang:2009wd}, but we spare the reader the derivation of these constraints as the detailed general form of $P^k_{12}$ for Coulomb-branch amplitudes is beyond the scope of this paper. In the following, we will only need the fact that that any SUSY invariant can be decomposed into terms of the form~(\ref{NkMHVSU2})
with $P^k_{12}$ of homogenous degree. It follows from this that $SU(2)$ superamplitudes can be organized into an MHV band with $\eta$ degrees $2,4,6$; an NMHV band with $\eta$-degrees $4,6,8$; an N$^2$MHV band with $\eta$-degrees $6,8,10$; etc. The different N$^k$MHV bands are thus all of the same ``width'' in $\eta$-degree and overlap: for example, the NMHV
band contributes to amplitudes that are MHV in the massless limit.

We can assign a different N$^k$MHV level to both $SU(2)$ subsectors, according to the structure~(\ref{NkMHVSU2}). From this perspective, the MHV-band superamplitude~(\ref{AMHV}) is MHV$\times$MHV: it is MHV in both $SU(2)$ sectors. More generally, the N$^k$MHV$\times$ N$^{k'}$MHV-band  superamplitude  $\A_n^{k,k'}$ 
takes the schematic form
\begin{equation}
 \A^{k,k'}_n~=~\sum P_{12}^{k}P_{34}^{k'} \times\massA_{12}\times  \massA_{34}\,.
\end{equation}
Each term in $\A^{k,k'}_n$ is of degree $2(k+1)$, $2(k+2)$, or $2(k+3)$ in $\eta_{i1}$, $\eta_{i2}$ and of degree $2(k'+1)$, $2(k'+2)$, or $2(k'+3)$ in $\eta_{i3}$, $\eta_{i4}$. This structure is illustrated in
figure~\ref{figsuperAmps}
in the introduction.

The all-$n$ amplitude~(\ref{WWphis})
of the maximally $SU(4)$ violating sector fully determines the MHV$\!\times\!\bMHV$ sector superamplitude
\begin{equation}
    \A_n^\text{MHV$\times\overline{\text{MHV}}$-band} 
    ~=~
    \frac{
    \<1^\perp2^\perp\>[1^\perp2^\perp]
    }{x_{13}^4\prod_{i=4}^{n}(x_{2 i}^2+m^2)}
    \times \massA_{12}\times \overline{\massA\,}_{34}\,,
\end{equation}
where $\overline{\massA\!}^{34}$ is the conventional parity conjugate of the invariant $\massA_{34}$, \ie it arises from $\massA_{34}$ by the map $[ij]\leftrightarrow \<ji\>$ and $\eta_{ia}\to \bar\eta_{ia}$ for $a=3,4$\,.\footnote{Note that this representation of the superamplitude in terms of $\eta_{i1},\eta_{i2}$ and $\bar\eta_{i3},\bar \eta_{i4}$ is distinct from the non-chiral representation of~\cite{Boels:2010mj} in terms of  $\eta_{i1},\bar\eta_{i2}$ and $\eta_{i3},\bar \eta_{i4}$.}

The MHV and MHV$\!\times\!\bMHV$ band are all that is needed to fully determine the $4$- and $5$-point superamplitudes. Explicitly, for the case of two adjacent massive lines,
\begin{equation}
\boxed{
\begin{split}
  \A_4~&=~
    -
    \frac{[1^\perp2^\perp]^2[34]}{\<34\>x_{13}^4(x_{2 3}^2+m^2)}
    \times \massA_{12}\times \massA_{34}\\[1ex]
    \A_5~&=~
    -
    \frac{\<1^\perp2^\perp\>[1^\perp2^\perp]}{x_{13}^4(x_{2 4}^2+m^2)(x_{2 5}^2+m^2)}\Biggl[\,\,
    \frac{[1^\perp2^\perp][3|m^2\!-\!x_{42}x_{25}|5]}{\<1^\perp 2^\perp\>\<34\>\<45\>}\times\massA_{12}\times\massA_{34}
    \,-\,\massA_{12}\times \overline{\massA\,}_{34}\,-\,\overline{\massA\,}_{12}\times \massA_{34}\\[.5ex]
  &   \hskip5.5cm+\frac{\<1^\perp2^\perp\>\<3|m^2\!-\!x_{42}x_{25}|5\>}{[1^\perp 2^\perp][34][45]}\times\overline{\massA\,}_{12}\times\overline{\massA\,}_{34}\,\,
    \Biggr]\,.
\end{split}
}
\end{equation}
We see that even entire superamplitudes that encapsulate all sectors can be reasonably simple on the Coulomb branch.

\subsection{Simplified superamplitudes for a smart choice of $q$-frame}
\label{s:smart}
The vector $q$ carries no physical significance in itself --- it is a reference vector that we choose to decompose amplitudes into helicity amplitudes. It should thus be thought of as a choice of basis. 
Picking the reference vector $q$ to be identical for all lines already leads to one significant simplification:
amplitudes with all-plus or all-minus helicity vector bosons vanish in this case \cite{Schwinn:2006ca}! This would not have been the case if we had picked distinct $q_i$ for each line. Indeed,
\begin{equation}
    \<W^-W^-g^-\>~=~~\frac{\<1^\perp 2^\perp\>[q_1 q_2] \<3^\perp|2|q_3]}{[1^\perp q_1][2^\perp q_2][3^\perp q_3]}~+~\text{cyclic}\,,
\end{equation}
vanishes  when we take all $q_i$ to be the same. In the superamplitude structure, this simplification manifests itself in the absence of $\eta^0$ contributions.

Is there a preferred choice of $q$ that simplifies the (super)amplitudes even further?
One natural condition to impose is that the perp'ed momenta satisfy momentum conservation among themselves,
\begin{equation}\label{q0}
    \sum p_i^\perp~=~0 \qquad \text{ for some }q=q_0\,.
\end{equation}
Note that this condition is {\em not} incompatible with $q\cdot p_i\neq 0$ for all $i$, which must be satisfied for $q$ to define a non-singular choice of
basis.
For the case of 2 massive lines $m_1=-m_2=m$, the condition implies
\begin{equation}
\label{q0cond}
    \frac{m^2}{2q_0\cdot p_1}\,q_0^\mu+\frac{m^2}{2q_0\cdot p_2}\,q_0^\mu~=~0\,\qquad \Rightarrow  \qquad q_0\cdot (p_1+p_2)~=~0\,,
\end{equation}
as is obvious from~(\ref{decpi}). The condition~(\ref{q0}) thus becomes a linear orthogonality condition on $q$.
As long as $p_1$ and $p_2$ are not collinear,
this condition can be satisfied  for some regular choice of $q$-basis.

With this choice of $q$, a number of
 striking
simplifications occur. The kinematic $K_4'$ in~(\ref{2mKs}) vanishes,
so this eliminates
the most cumbersome term in the superamplitude $\massA_{12}$ in~(\ref{massA12}). The remaining $K$'s organize themselves to satisfy $K_4=K_2K_6$ for $q=q_0$. The invariant $\massA_{12}$ then takes on an intriguing factorized form:
\begin{equation}
  \massA_{12}~=~\biggl[\delta^{(4)}\!\bigl( |i^\perp\>\eta_{ia}\bigr)+\frac{m\<1^\perp2^\perp\>}{\<q1^\perp\>\<q2^\perp\>}\,\delta^{(2)}\!\bigl( \<qi^\perp\>\eta_{ia}\bigr)\biggr]\times
  \biggl[1-\frac{[1^\perp q][2^\perp q]}{m[1^\perp 2^\perp]}\,\delta^{(2)}\!\bigl( \mu_i\eta_{ia}\bigr)\biggr]
  \qquad\text{ for }q=q_0\,.
\end{equation}
The entire MHV-band superamplitude, for example, then simply reads
\begin{equation}
\boxed{
\begin{split}\label{MHVspecialq}
    \A_n^{\rm MHV-band}
    =~&
    \frac{
    -
    [3|\prod_{i=4}^{n-1}[m^2\!-\!x_{i2}x_{2,i+1}]|n]}{\<1^\perp2^\perp\>^2\<34\>\<45\>\cdots \<n\!-\!1,n\>\prod_{i=4}^{n}(x_{2 i}^2\!+\!m^2)}\\
&\times\biggl[\delta^{(4)}\!\bigl( |i^\perp\>\eta_{ia}\bigr)\!+\!\frac{m\<1^\perp2^\perp\>}{\<q1^\perp\>\<q2^\perp\>}\,\delta^{(2)}\!\bigl( \<qi^\perp\>\eta_{ia}\bigr)\biggr]^2\!\times
  \biggl[1-\frac{[1^\perp q][2^\perp q]}{m[1^\perp 2^\perp]}\,\delta^{(2)}\!\bigl( \mu_i\eta_{ia}\bigr)\biggr]^2
  \quad\text{ for }q=q_0\,,
\end{split}
}
\end{equation}
where the squares should be understood as a product of two factors corresponding to the two $SU(2)$'s.
Here we also used the relation $x_{13}^2=2 p_1^\perp\cdot p_2^\perp$ which holds for this special choice of $q$.

A similar factorization also occurs
 for
superamplitudes with arbitrarily many massive legs
when \reef{q0}  is imposed;
this will be discussed in section~\ref{supermultimassive} below.

\subsection{3-point superamplitudes}
\label{s:3ptsuperA}
So far we have studied superamplitudes with $n>3$. However,
the 3-point superamplitudes on the Coulomb-branch 
also 
take a very simple form. As the explicit
$\<W \,\overline{W}\, g\>$
amplitudes~(\ref{3point}) suggest,
the massive generalizations of the conventional MHV and $\bMHV$ sectors follow from the conventional Parke-Taylor superamplitude by replacing conventional spinors $|i\>$, $|i]$ by perp'ed spinors $|i^\perp\>$, $|i^\perp]$:
\begin{equation}\label{3pointsuper}
    \A^{\rm MHV}_3~=~\frac{\delta^{(8)}\big(|i^\perp\>\eta_{ia}\big)}{\<1^\perp2^\perp\>\<2^\perp3^\perp\>\<3^\perp 1^\perp\>}\,,\qquad
    \A^{\bMHV}_3~=~\frac{\delta^{(4)}\big([1^\perp2^\perp]\eta_{3a}\!+\!\text{cyc}\big)}{[1^\perp2^\perp][2^\perp3^\perp][3^\perp 1^\perp]}\,.
\end{equation}

 The $SU(4)$-violating MHV$\!\times\!\bMHV$ sectors are completely determined  by SUSY from the MHV and anti-MHV sectors. Their Grassmann structure within each $SU(2)$ sector is either of the MHV or $\bMHV$ type --- it remains to determine the overall kinematic prefactor. For the case of two massive lines $m_1=-m_2=m$, the MHV$\!\times\!\bMHV$ superamplitude is given by
\begin{equation}
\begin{split}
    \A^{\rm MHV\!\times\!\bMHV}_3~&=~\frac{m\<q|3|q]^2}{\<q|1|q]\<q|2|q][1^\perp2^\perp]\<1^\perp 2^\perp\>}\times\delta_{12}^{(4)}\big(|i^\perp\>\eta_{ia}\big)\times\delta_{34}^{(2)}\Big([1^\perp2^\perp]\eta_{3a}\!+\!\text{cyc}\Big)\,.
\end{split}
\end{equation}
Here the subscripts on $\delta$ signify the $SU(2)$ sector that the Grassmann delta-function lives in.
The parity-conjugate superamplitude $\A^{\rm \bMHV\!\times\!MHV}_3$ is identical, except that the two $SU(2)$ sectors are exchanged: $\delta_{12}\leftrightarrow\delta_{34}$.

There is another form of the MHV$\!\times\!\bMHV$ sector with a slightly different packaging of the $\eta$-structure,
\begin{equation}
\begin{split}
    \A^{\rm MHV\!\times\!\bMHV}_3~&=~\frac{m}{\<q|1|q]\<q|2|q]}\times
    \prod_{a=1}^2\!\!\Big([1^\perp q]\eta_{2a}\eta_{3a}+\text{cyc}\Big)\times\delta_{34}^{(2)}\big(\<qi^\perp\>\eta_{ia}\big)\,.
\end{split}
\end{equation}
When all three lines are massive,\footnote{For this to happen, the gauge group must be broken into a product of at least 3 $SU(N_i)$'s.} subject to the  6d momentum conservation constraint $m_1+m_2+m_3=0$, this superamplitude generalizes to:
\begin{equation}
\begin{split}\label{SU4viol3pt}
    \A^{\rm MHV\!\times\!\bMHV}_3~&=~\frac{m_2\<q|1|q]-m_1\<q|2|q]}{\<q|1|q]\<q|2|q]\<q|3|q]}\times
    \prod_{a=1}^2\!\!\Big([1^\perp q]\eta_{2a}\eta_{3a}+\text{cyc}\Big)\times\delta_{34}^{(2)}\big(\<qi^\perp\>\eta_{ia}\big)\,.
\end{split}
\end{equation}
 This form is useful for our discussion of the CSW-expansion for Coulomb-branch amplitudes in section \ref{sec:CSW}.

%%%%%%%%%%%%%%%%
%%%%%%%%%%%%%%%%
\subsection{Single-soft limits and matching at the superamplitude level}
In section \ref{s:soft} we saw examples of how the leading small-mass limit of massive amplitudes could be obtained from soft scalar  limits of massless amplitudes at the origin of moduli-space. We now take this one step further and show how the matching can be promoted to superamplitudes.

As the starting point, consider the $n$-point MHV-band superamplitude with two massive external particles on lines 1 and 2, $m_1 = -m_2=m$. The MHV-band superamplitude has a clear hierarchy of mass-powers, and expanding in small $m$ up to order $\eta^6$, we find that the superamplitude~(\ref{AMHV}) is simply\footnote{To leading order, the $\eta^6$ pieces in the MHV-band superamplitude correspond to actual amplitudes --- the $\eta^6$ pieces in the MHV\!$\times$\!NMHV and NMHV\!$\times$\!MHV bands only contribute to order $O(m^2)$.}
\bea
  \nonumber
  {\mathcal{A}}_n^\text{MHV-band}
  \!\!&\!\!\!=\!\!\!&\!\!
  \frac{1}{\<1^\perp2^\perp\> \cdots \<n1^\perp\>}\times \frac{m\<1^\perp2^\perp\>}{\<q1^\perp\>\<q2^\perp\>}\times
  \bigg\{\delta_{12}^{(4)}\!\bigl(|i^\perp\>\eta_{ia}\bigr)\,\delta_{34}^{(2)}\bigl( \<qi^\perp\>\eta_{ia}\bigr)
  +\delta_{12}^{(2)}\bigl( \<qi^\perp\>\eta_{ia}\bigr)\,\delta_{34}^{(4)}\!\bigl(|i^\perp\>\eta_{ia}\bigr)\bigg\}\\[1.25ex]
   &&~~~~~~~+O(m^2)+O(\eta^8)\,.
  \label{superAn}
\eea
Here, the factor of $m\<1^\perp2^\perp\>/\<q1^\perp\>\<q2^\perp\>$ is simply the small-mass expansion of the kinematic factor $K_2$ in~(\ref{2mKs}).
The states associated with the amplitudes in~(\ref{superAn})  contain two pairs of indices $a=1,2$ and one pair of $a=3,4$ (or the reverse). Clearly this violates $SU(4)$. An example of an amplitude in this sector is  $\bigl\<W^- \barW^+\phi^{34}g^+\cdots g^+\bigr\>$. Projecting out this particular amplitude from \reef{superAn} produces the leading order expression that we stated in \reef{WWphi34}.

Let us now compare the massive superamplitude \reef{superAn} with the single-soft scalar limit of the $(n+1)$-point MHV superamplitude at the origin of moduli space:
\bea
   \mathcal{A}_{n+1}^\text{MHV}(1, \eps q, 2,3, \dots, n) &=& \frac{1}{\<1,\epsilon q\>\<\epsilon q,2\>\<23\> \cdots \<n1\>}\,
  \d^{(8)}\Big(|\epsilon q\>\eta_{i,\epsilon q}+ \sum_{i=1}^n  |i\> \eta_{ia} \Big)   \,.
\eea
We now follow the same steps as in section~\ref{3.1}:
On the line with momentum $\eps q$, we project out the linear combination of scalars $\phi^{12}+\phi^{34}$ corresponding to the vev, and take the soft scalar limit $\eps \to 0$ as discussed above. We find
\begin{equation}
\begin{split}
   &\mathcal{A}_{n+1}^\text{MHV}(1, \phi_{\eps q}^{12}, 2 ,3, \dots, n)+\mathcal{A}_{n+1}^\text{MHV}(1, \phi_{\eps q}^{34}, 2 ,3, \dots, n)
   \\&=
   \frac{1}{ \<1q\>\<q2\>\<23\> \cdots \<n1\>}
  \biggl\{
  \d^{(4)}_{12}\Big( \sum_{i=1}^n  |i\> \eta_{ia}\Big)\sum_{i,j}  \<q i\>\<q j\> \eta_{i3}\eta_{i4}
  +\d^{(4)}_{34}\Big( \sum_{i=1}^n  |i\> \eta_{ia} \Big)\sum_{i,j}  \<q i\>\<q j\> \eta_{i1}\eta_{i2}
  \biggr\} \,.
  \label{SSLsuper}
\end{split}
\end{equation}
Note that the sum in the delta-function does not include 
line $q$ anymore.
Comparing this to~(\ref{superAn}), we have confirmed
\begin{equation}
    {\mathcal{A}}_n^\text{MHV-band}=-m\Bigl[\mathcal{A}_{n+1}^\text{MHV}(1, \phi_{\eps q}^{12}, 2 ,3, \dots, n)+\mathcal{A}_{n+1}^\text{MHV}(1, \phi_{\eps q}^{34}, 2 ,3, \dots, n)\Bigr]+O(\eta^8)+O(m^2)
\end{equation}
at the level of superamplitudes.

\subsection{Multi-soft limits of N$^k$MHV superamplitudes}
\label{s:multisoft}

We now present a simple expression for multi-soft scalar limits of the massless N$^k$MHV superamplitude in the case where all soft scalars are adjacent and identical, say $\phi^{12}$.
As explained in section~\ref{s:genms}, the identical-scalar soft limits are indeed all that is needed to determine the leading order terms in amplitudes that are $O(m^3)$ or higher.
In particular, this computation allows us to match to the leading-mass term $O(m^{n-2})$ of the $n$-point maximally-$SU(4)$-violating amplitude
$\<W^- \overline{W}^+ \phi^{34} \dots \phi^{34}\>$
of (\ref{WWphis}) from a $(n\!-\!2)$-fold soft scalar limit of the $(2n\!-\!2)$-point N$^{n-3}$MHV amplitude
$\<g^- \phi^{12} \dots \phi^{12} g^+ \phi^{34} \dots \phi^{34}\>$.

As a warmup, let us first consider the double soft limits of the MHV and NMHV superamplitudes.
We take lines  $1$ and $N$ to be the scalars $\phi^{12}$; practically this is done by applying Grassmann derivatives
$\partial_1^1\partial_1^2 \partial_N^1\partial_N^2$ and subsequently setting $\eta_{1a},\eta_{Na} \to 0$ for $a=3,4$.
At MHV level,
we have
\begin{equation}
\label{2softMHV}
\mathcal{A}_{N,\,\phi^{12}_1\phi^{12}_N}^\mathrm{MHV}
~=~ - \frac{\delta^{(4)}_{34} \big(\sum_{i=2}^{N-1} |i \rangle \eta_i \big)  }{\langle 2 3 \rangle \langle 3 4 \rangle ... \langle N\!-\!2 , N\!-\!1 \rangle}
\times \frac{\langle N 1 \rangle}{\langle N\!-\!1 , N \rangle \langle 1 2 \rangle}.
\end{equation}
Note that the expression on the RHS is valid even without taking $p_1$ and $p_N$ soft; we will need that later.
To compare with our massive amplitudes with all reference vectors equal, we take a collinear limit $p_1,p_N \to q$ before taking the momenta soft.
Due to the $\<N1\>$ numerator-factor in \reef{2softMHV}, the MHV contribution vanishes in the collinear limit, so $\mathcal{A}_{N,\,\text{2-soft}}^\mathrm{MHV}=0$. This agrees with our previous results, since the
Coulomb-branch superamplitudes do not\footnote{
If we had not insisted on all-$q_i$ reference vectors equal, such terms could appear in the massive superamplitude,  and indeed \reef{2softMHV} with $p_1 \ne p_N$ indicates that they could be matched by  soft-scalar limits. We will not pursue this here.
}
 contain
 terms of zeroth order in the $\eta$'s of one $SU(2)$ sector.

Now consider the double soft limit of the NMHV superamplitude, again with soft $\phi^{12}$ scalars on lines $1$ and $N$.
The superamplitude \cite{Drummond:2008bq} can be written
\begin{equation}
\mathcal{A}_N^\mathrm{NMHV} ~=~ \mathcal{A}_N^\mathrm{MHV}\!\!\! \sum_{2 \le a_1, b_1 \le N-1}\!\!\!\! R_{N;a_1 b_1}\,,
\end{equation}
where
 $a_1 +1 < b_1$ in the sum and $R_{N;a_1 b_1}$ is the dual superconformal invariant given by
\begin{align}
R_{N;a_1 b_1} \,=\, & \frac{\langle a_1 , a_1 \!-\!1 \rangle \langle b_1, b_1 \!-\!1 \rangle \delta^{(4)}\big(
\langle N | x_{N a_1} x_{a_1 b_1} | \theta_{b_1 N} \rangle + \langle N | x_{N b_1} x_{b_1 a_1} | \theta_{a_1 N} \rangle \big)
}{x_{a_1 b_1}^2 \langle N | x_{N a_1} x_{a_1 b_1} | b_1 \rangle \langle N | x_{N a_1} x_{a_1 b_1} | b_1 \!-\! 1\rangle \langle N | x_{N b_1} x_{b_1 a_1} | a_1\rangle \langle N | x_{N b_1} x _{b_1 a_1} | a_1 \!-\! 1 \rangle}
\,.
\end{align}
Here we have defined $ \theta_{uv} = \sum_{i=u}^{v-1} |i \rangle \eta_i$ for $u < v$, $\theta_{uv} = - \theta_{vu}$ for $v < u$.

The $R$-invariant $R_{N;a_1 b_1}$ has no explicit dependence on
$\eta_1, \eta_N$, so inserting
$\phi^{12}$
on those lines
gives
\begin{equation}
   \mathcal{A}_{N,\,\phi^{12}_1\phi^{12}_N}^\mathrm{NMHV}
    ~=~\mathcal{A}_{N,\,\phi^{12}_1\phi^{12}_N}^\mathrm{MHV}
     ~~\times \!\!\!\sum_{2 \le a_1, b_1 \le N-1}\!\!\!\! R_{N;ij}\,.
    \label{NMHV2softA}
\end{equation}
Since the prefactor
$ \mathcal{A}_{N,\,\phi^{12}_1\phi^{12}_N}^\mathrm{MHV}$
vanishes in the collinear limit, any finite contribution must come from $R$-invariants with simple poles in the sum. Note that $R_{N;2 b_1}$ has denominator terms such as $\langle N | x_{N 2} \!\cdots \!=\! \langle N | p_1 \!\cdots \!=\! \<N1\>[1|\!\cdots$, and indeed these are the only $R$-invariants in the sum \reef{NMHV2softA} with such poles in the collinear limit. However, in one case, namely $b_1=N\!-\!1$, the denominator factors $\<N1\>$ are cancelled by contributions from $\delta^{(4)}$, and a detailed calculation shows that this particular contribution vanishes like $[N1]$ in the collinear limit. The remaining $R_{N;2 b_1}$ give finite results, and we find  the NMHV double-soft limit to be
\begin{equation}
\mathcal{A}_{N,\,\text{2-soft}}^\mathrm{NMHV}\,=\,
- \frac{\delta^{(4)}_{34} \big(\sum_{i=2}^{N-1}\! |i \rangle \eta_i \big)  }
{\langle 2 3 \rangle \langle 3 4 \rangle \ldots \langle N\!-\!1 , q \rangle \langle q 2 \rangle} \!
 \sum_{3 < b_1 < N\!-\!1}
\!\!\!\! \frac{\langle b_1 \!-\!1 , b_1 \rangle
~ \delta^{(2)}_{34} \big( [q| x_{2 b_1} |
\sum_{i=b_1}^{N-1} |i\> \eta_{i}
\rangle  \big)
~ \delta^{(2)}_{12} \big(\sum_{i=2}^{N-1}\! \langle q  i \rangle \eta_i \big) }
{
P_{b_1 \dots N-1}^2
[q| x_{2 b_1} | b_1 \rangle [q| x_{2 b_1} | b_1\! -\! 1\rangle }\,.
    \label{NMHV2softB}
\end{equation}

The sum in \reef{NMHV2softB} is empty for $N=5$, so let us
consider the simplest non-vanishing case, $N=6$.
For a proper association with a Coulomb-branch amplitude with massive lines $1$ and $2$, we relabel momenta $\{q,2,3,4,5,q\}\to\{q,2,3,4,1,q\}$ to find
\begin{equation} \label{6ds}
\mathcal{A}_\text{6,\,2-soft}^\mathrm{NMHV}~=~
 \frac{\delta^{(4)}_{34} \big(\sum_{i=1}^{4} |i \rangle \eta_i \big)
 ~ \delta^{(2)}_{34} \big([ 4q]  \eta_1 - [ 1q]  \eta_4  \big)
 ~ \delta^{(2)}_{12} \big(\sum_{i=1}^{4} \langle q  i \rangle \eta_i \big) }{ [14] \langle 14 \rangle [1q ]  \langle q1\rangle [2q] \langle q 2 \rangle \langle 2 3 \rangle^2}\,.
\end{equation}

As an example, consider the component amplitude
 $\< g^- \phi^{12}_{\epsilon q}\phi^{12}_{\epsilon q} g^+ \phi^{34} \phi^{34}\>$. It can be easily extracted from~(\ref{6ds}), giving
 \begin{equation}
\lim_{\epsilon\to 0}\bigl\< g^-_1 \,\phi^{12}_{\epsilon q}\,\phi^{12}_{\epsilon q} \,g^+_2\, \phi^{34}_3\, \phi^{34}_4\bigr\>~=~
- \frac{ \langle q1 \rangle [2q]  }{ [23] \langle 23 \rangle \langle q2 \rangle [1q] }\,.
 \end{equation}
This is precisely the soft-limit given in~(\ref{6ptsoft}) that we used to determine the leading term in $\bigl\< W^-\barW^+\, \phi^{34}\, \phi^{34}\bigr\>$.
Thus we have successfully made contract with the soft-scalar limits at the NMHV level.

For the general $N$-point
N$^k$MHV superamplitude, we adopt the notation of \cite{Drummond:2008cr}; in particular, we use the  diagrammatic expansion given in Figure 4 of \cite{Drummond:2008cr}.
We restrict ourselves to the case where all the scalars are adjacent: we choose them to lie on lines $N, 1, ... s-1$ for the $s$-scalar soft limit of the $N$-point N$^k$MHV massless superamplitude. The resulting soft-limit thus corresponds to the leading $O(m^s)$ contribution in an $n$-point amplitude with $n=N-s$.
It turns out that the only
non-vanishing contribution to the
$s$-scalar soft limit,
with soft $\phi^{12}$ scalars on lines $N, 1,\ldots,s-1$, is given by the left-most branch of the expansion, with $a_i = i+1$ for $1 \le i < s$. Each new level\footnote{
Each term in an N$^k$MHV amplitude contains $k$ factors of $R$-invariants.
We denote the first, NMHV-like factor as ``level 1'', the N$^2$MHV-like factor as ``level 2'', etc.}
 of the expansion
contributes a factor of the form
\begin{equation} \label{Rn}
R_{N;b_1 2; b_2 3; ... ; b_{i-1} i; i+1 \, b_i}
~\rightarrow~
 \frac{  \langle N  i \rangle \langle i, \, i +1\rangle}{ \langle N, \,  i+1 \rangle}\times  \frac{ \langle  b_i -1 , \, b_i \rangle  \delta^{(2)}_{34} \big( [i| x_{i+1 \, b_i} | \theta_{b_i \, i+1} \rangle \big)}{  x_{i+1 \, b_i}^2  [i| x_{i+1 \, b_i} | b_i - 1 \rangle  [i| x_{i+1 \, b_i} | b_i  \rangle  }. \end{equation}
In the same way that the $j=N-1$ term vanished in the NMHV case, the ``boundary'' terms where $b_i = b_{i+1}$ vanish in the more general case; in order to be able to continue down to level $s-1$ of the expansion, we require $s+1 = a_{s-1} +1 < b_{s-1} < b_{s-2} < ... < b_1 < N$. Terms on other branches of the expansion, or not satisfying $a_i = i+1$, can be shown to vanish
on the support of the delta functions when the $s$ lines are taken
collinear and the shared momentum is taken soft.

Note that the $\langle i, \, i\!+\!1\rangle$ factor in each term
in~(\ref{Rn})
cancels a pole from the MHV prefactor, and the $\langle N \, i \rangle / \langle N , i\!+\!1 \rangle$ factor telescopes.
The product of these factors with the leading MHV and NMHV-like prefactors   are  manifestly finite and non-zero, and we find
\begin{equation}\label{TS}
\boxed{
 \begin{split}
    \mathcal{A}_{N,\,s\text{-soft}}^{\mathrm{N}^k\mathrm{MHV}}
 & = - \frac{\delta^{(4)}_{34} \big( \sum_{i=s}^{N-1} |i \rangle \eta_i \big)  \delta^{(2)}_{12}  \big(\sum_{i=s}^{N-1} \langle q i \rangle \eta_i \big) }{\langle s , s\!+\!1 \rangle ... \langle N\!-\!2 , N\!-\!1 \rangle \langle N\!-\!1 , q \rangle \langle q \, s \rangle}  \\
 &~~~~  \times  \!\!\!\sum_{s+1 < b_{s-1} < ... < b_1 < N\!-\!1}\,  \prod_{i=1}^{s-1} \frac{ \langle b_i\!-\!1, \, b_i \rangle \delta^{(2)}_{34} \big( [q| x_{N \, b_i} | \theta_{b_i \, N} \rangle \big) }{  x_{b_i \, s}^2 [q| x_{N \, b_i} | b_i \rangle  [q| x_{N \, b_i} | b_i \!-\!1\rangle}  \\
 &  \hskip4cm\times \text{remaining product of $(k-s+1)$ $R$-invariants}\,.~~~~
 \end{split}
}
\end{equation}
 Here, we have plugged in the soft momentum $q$ on lines $N,1,\ldots,s-1$, but we have {\em not} relabeled the lines $s,\ldots,N-1$.
 If $k > s-1$, the diagrammatic expansion continues down to level $k$, and is a sum over all terms descending from the vertices on the leftmost branch at level $s-1$, with each vertex characterized by the choice of previous
 $b_i$-indices
 $\left\{ b_1, b_2, \ldots, b_{s-2}, b_{s-1} \right\}$. Each such term is a product of $(k-s+1)$ $R$-invariants.
In all such descendant factors, we can set $\eta_i\to0$ for $i=N,1, \ldots,s-1$.\footnote{A further simplification can be applied to the descendant $R$-invariants: The spinor  $\langle \xi| = \langle N | x_{N b_1} \!\cdots\!x_{s b_s}\!\cdots\!  x_{b_r a_r}$ defined in \cite{Drummond:2008cr} can be replaced by $ \langle \xi| \rightarrow \langle q | x_{s b_s}\!\cdots\! x_{b_r a_r}.$}

\vspace{2mm}
\noindent {\bf Example: Matching to maximally-$SU(4)$-violating amplitudes}\\
As a concrete example, consider the leading term in the maximally-$SU(4)$-violating amplitude of~(\ref{WWphis}), $\< W^- \barW^+\, \phi^{34} \cdots \phi^{34}\>_n$. Its leading behavior is $O(m^{n-2})$:
\begin{equation}\label{leadingRviol}
        \bigl\< W^-_1 \barW^+_2\, \phi^{34}_3\, \phi^{34}_4 \cdots \phi^{34}_n\bigr\>
   ~=~-m^{n-2}\frac{\<1^\perp|q|2^\perp]}{\<2^\perp|q|1^\perp]\prod_{i=4}^{n}x_{2 i}^2}~+~O(m^{n})\,.
\end{equation}

According to the proposal of section~\ref{s:genms}, this leading behavior can be obtained from the soft limit~(\ref{TS})  with
$k=n-3$, $s=k+1$ and $N=n+k+1$ ($= n+s = 2s+2$). This is the simplest non-vanishing
example because no remaining product of $R$-invariants appears in~(\ref{TS}) in this case, and there is exactly one allowed choice for
the $b_i$'s: $\{b_1, b_2, ..., b_{s-2}, b_{s-1} \} = \{ N-2, N-3, ..., s+3, s+2\}$, i.e.  $b_i = N-1-i$ for $i < s$.
Projecting out a positive helicity gluon on  line $s$,
a negative helicity gluon on line $N-1$ and $\phi^{34}$ scalars on the other lines is done easily when the  $\delta^{(4)}$-function is used to rewrite $|\theta_{b_i\,N}\> = - |\theta_{s b_i}\>$. We obtain the soft-limit component amplitude,
\begin{equation}
\lim_{\epsilon\to 0}\bigl\<\phi^{12}_{\epsilon q}\cdots \phi^{12}_{\epsilon q}\, g^+_s\,\phi^{34}_{s+1}\cdots \phi^{34}_{N-2}\,g^-_{N-1}\bigr\>~=~
 \frac{ \langle N\!-\!2, \, N\!-\!1 \rangle \langle N\!-\!1, \, q \rangle} {\langle s, \, s\!+\!1 \rangle   \langle q, \, s \rangle\,\prod_{i=1}^{s-1} x_{s+1+i, \, s}^2}
 ~\prod_{i=1}^{s-1} \frac{    [q| x_{N, \, s+1+i} | s\!+\!i \rangle} {[q| x_{N, \, s+1+i} | s\!+\!1\!+\!i \rangle }. \end{equation}
Using momentum conservation in the multi-soft limit, we can write,
\[\prod_{i=1}^{s-1} \frac{ [q | x_{N, \, s+1+i} | s\!+\!i \rangle}{[q | x_{N, \, s+1+i} | s\!+\!1\!+\!i \rangle} =  \frac{[q|x_{N \, s+2} | s\!+\!1 \rangle}{[q|x_{N, \, N-1} | N\!-\!2 \rangle} = \frac{ [q s] \langle s, \, s\!+\!1 \rangle}{[q, \, N-1] \langle N\!-\!2, \, N\!-\!1 \rangle}. \]
Relabeling
$\{s,s+1,\ldots,N-1\}\to\{2,3,\ldots,n,1\}$
gives the final result,
\begin{equation}
\lim_{\epsilon\to 0}\bigl\<g^-_{1}\,\phi^{12}_{\epsilon q}\cdots \phi^{12}_{\epsilon q}\, g^+_2\,\phi^{34}_{3}\cdots \phi^{34}_{n}\bigr\>~=~
- \frac{ \langle 1 \, q \rangle [q \, 2]}{\langle 2 \, q \rangle[ q \, 1]  } \frac{1}{\prod_{i=4}^{n} x_{2i}^2}\,. \end{equation}
This matches precisely to the leading term in the massive amplitude~(\ref{leadingRviol}) for all $n$, consistent with the proposed relation~(\ref{proposalmk}) of section~\ref{s:genms}.

\subsection{The structure of superamplitudes with generic massive lines}
\label{supermultimassive}

In the above discussion we focused on amplitudes with only two massive lines 
adjacent to each other in the trace-structure. We argued that the corresponding superamplitudes must be proportional to $\massA_{12}\times \massA_{34}$, defined in~(\ref{massA12}), with the kinematic coefficient $K_i$ given in~(\ref{2mKs}).
$\massA_{12}\times \massA_{34}$ is the Coulomb-branch generalization of the conventional  $\delta^{(8)}(|i\>\eta_{ia})$ in the massless case.

It turns out that there is also a simple and concise expression for the generalization of $\delta^{(8)}(|i\>\eta_{ia})$ to the case with arbitrarily many massive lines. In this case $\massA_{12}$ is still given by~(\ref{massA12}).
To write down the general kinetic factors $K_2$, $K_4$, $K_4'$ and $K_6$, however, it is convenient to  introduce a bit of notation.

We define  $K^\perp$ as the constant satisfying
\begin{equation}
    \sum_i |i^\perp\>[i^\perp| = K^\perp\, |q\>[q|\,.
\end{equation}
From~(\ref{decpi}),
it is obvious that $K^\perp$ can be written as
\begin{equation}
    K^\perp~=~\sum_i\frac{m_i^2}{2q\cdot p_i}~=~-\sum_i\bar\mu_i\mu_i\,.
\end{equation}
Next, we note that
\begin{equation}
    \sum_i\mu_i[i^\perp|\propto [q|\,,\qquad \sum_i\bar\mu_i|i^\perp\>\propto |q\>\,.
\end{equation}
This follows immediately from the 6d momentum conservation constraint $\sum_i m_i=0$, as can be seen from contracting the left-hand-sides with $|q]$ and $\<q|$, respectively. We now denote the kinematic proportionality constants by  $K^\parallel$ and $\bar K^\parallel$, respectively:
\begin{equation}
     \qquad\sum_i\mu_i[i^\perp|~=~K^\parallel\, [q|\,,\qquad
    \sum_i\bar\mu_i|i^\perp\>~=~\bar K^\parallel\, |q\>\,.
\end{equation}
Explicit expressions for $K^\parallel$ and $\bar K^\parallel$ can be obtained by contracting the above equations with arbitrary spinors $\neq |q],\<q|$\,.

Using $K^\perp$, $K^\parallel$ and $\bar K^\parallel$, it is now easy to write down explicit expressions for the kinematic constants $K_i$ that appear in $\massA_{12}$:
\begin{equation}
\label{K246genM}
  K_2~=~\frac{(K^\perp)^2}{K^\parallel}+\bar K^\parallel\,,\qquad
  K_4~=~\frac{\bar K^\parallel}{K^\parallel}\,,\qquad
  K_4'~=~
  \frac{K^\perp}{K^\parallel}\,,\qquad
  K_6~=~\frac{1}{K^\parallel}\,.
\end{equation}
It is straight-forward (albeit tedious) to verify that with these $K_i$, the superamplitude  $\massA_{12}$ is invariant under all supercharges~(\ref{massive2comp12}).

We note that the ``special choice'' of $q=q_0$ 
discussed above in the $2$-mass case has an arbitrary-mass generalization. We again demand
\begin{equation}
    \sum_ip_i^\perp~=~0\,.
\end{equation}
In the two-mass case, this  condition reduced to the orthogonality condition $q_0\cdot(p_1+p_2)=0$ given in~(\ref{q0cond}). For more than 2 massive lines, this condition is no longer linear and
it is
therefore
less practical to implement (though solutions $q_0$ are still guaranteed to exist). For $q=q_0$, we have $K^\perp=0$ and thus $K_4'=0$ and $K_4=K_2K_6$, just as in the 2-mass case.
With this special choice of $q$, the invariant $\massA_{12}$ takes the following simple factorized form:
\begin{equation}
  \massA_{12}~=~\Bigl[\delta^{(4)}\!\bigl( |i^\perp\>\eta_{ia}\bigr)+\bar K^\parallel\delta^{(2)}\!\bigl( \<qi^\perp\>\eta_{ia}\bigr)\Bigr]\times
  \Bigl[1+\frac{1}{K^\parallel}\delta^{(2)}\!\bigl( \mu_i\eta_{ia}\bigr)\Bigr]
  \qquad\text{ for }q=q_0\,.
\end{equation}
Due to its simplicity, this form provides a natural starting point for a comparison to the non-chiral formulation
that arises from a straight-forward dimensional reduction of the 6d superamplitudes of~\cite{Bern:2010qa,Brandhuber:2010mm,Dennen:2010dh}.

\subsection{4d-6d correspondence}
We have several times alluded to the connection between 6d massless amplitudes of $\cn=(1,1)$ SYM and the 4d $\cn=4$ SYM amplitudes on the Coulomb branch. In this section we briefly outline the map to go between the two descriptions.

The 4d masses are simply the extra dimensional momenta $p_4 \pm ip_5 = m_\pm$, and hence the external states of the on-shell amplitudes have to satisfy the  condition $\sum_{i=1}^n m_i = 0$. The 6d spinor-helicity formalism \cite{Cheung:2009dc} can easily be decomposed to 4d massive spinors, as for example the 6d spinor product (in our conventions)

\bea
\langle i_{\raisebox{-1.5pt}{$\scriptstyle a$}} | j_{\dot b} ]
&=&
 \left( \begin{array}{cc}
 - \< i^\perp j^\perp\>
 &  \mu_i [j^\perp q]
     -  \bar \mu_j  \<q i^\perp \>
 \\[2mm]
 -  \bar \mu_i  \<q j^\perp \>
 +  \mu_j [i^\perp q]
      &  [ i^\perp j^\perp ]  \end{array} \right)\,.
\eea

The 6d on-shell superspace reduces to a non-chiral representation of 4d on-shell superspace \cite{Hatsuda:2008pm,Bern:2010qa,Brandhuber:2010mm,Dennen:2010dh}; the map can be found in section 7.2 of \cite{Elvang:2011fx}. To recover the familiar chiral superfield formulation, a half-Fourier transform is carried out on the Grassmann variables. In the massless case, this allows us to trace which terms in the 6d superamplitudes give rise to the N$^k$MHV classification of the 4d superamplitudes. Similarly, it should be possible to extract the N$^{k}$MHV$\times$N$^{k'}$MHV-band structure for the massive amplitudes; this however is non-trivial, in part because only 3,4,5-point 6d superamplitudes are currently available. We have only checked the 4d-6d match of the massive MHV band for the $n=4$ case.

%%%%%%%%%%%%%%%%
%%%%%%%%%%%%%%%%
%%%%%%%%%%%%%%%%
\setcounter{equation}{0}
\section{CSW expansion for all MHV$\!\times\!$N$^k$MHV superamplitudes}\label{sec:CSW}
We now derive 
an expression for the tree-level
MHV$\!\times\!$N$^k$MHV superamplitudes 
 with arbitrarily many massive lines.
 The derivation is based on a CSW-type expansion of the superamplitude in terms of Coulomb-branch 3-point superamplitudes.

To illustrate the method, we consider the gauge boson amplitudes $\< X^-_1 X^+_2\,\cdots X_n^+\>$, where each state $X$ can
be a massive $W$ boson or a massless gluon $g$. For the case of only two massive $W$ bosons on lines $1$ and $2$, this amplitude was given above in~(\ref{WWalln}).
Here we let the $n$ vectors have arbitrary masses $m_i$, subject to $\sum m_i=0$.
Consider the holomorphic all-line shift\footnote{
Massless holomorphic all-line (super)shifts were introduced in~\cite{Elvang:2008vz,Kiermaier:2009yu} to prove the (super-)MHV vertex expansion in $\cn=4$ SYM (see also~\cite{Bullimore:2010dz}). They generalize the Risager 
3-line
shift~\cite{Risager:2005vk}, which was applied in $\cn=4$ SYM to derive MHV vertex expansions at the NMHV level~\cite{Bianchi:2008pu,Elvang:2008na}. The generalization of holomorphic all-line shifts to massive external lines was presented in~\cite{Cohen:2010mi}. }
\begin{equation}
\label{allline}
   |i^\perp\>~\to~|\hat{i}^\perp\> = |i^\perp\> + z\,c_i\, |q\> \, ,
   ~~~~~\text{with}~~~~\sum_i c_i |i^\perp] = 0 \,,
\end{equation}
in which $q$ is the \emph{same} reference spinor that we use in our massive spinor helicity formalism. It follows from the proof in \cite{Cohen:2010mi} that amplitudes $\< X^-_1 X^+_2\,\cdots X_n^+\>$ with $n>3$ fall off as $1/z$, or better, for large-$z$. The resulting all-line recursion relations will allow us to construct these $n$-point amplitudes from $\< X^-_1 X^+_2\, X_3^+\>$  as  a simple CSW-type $\overline{\text{MHV}}$ vertex expansion, as we now show.

As a simple warm-up, consider $n=4$. The all-line shift recursion relation in this case has two diagrams, corresponding to the $P_{12}$ and $P_{23}$ channel, and it gives
\begin{equation}
\begin{split}
    &\big\< X^-_1 X^+_2\, X_3^+\, X_4^+\big\>\\
    &=~
    \frac{[2^\perp P_{12} ]^3}{[1^\perp2^\perp][P_{12}1^\perp]}\!\times\!\frac{1}{P_{12}^2+m_{12}^2}\!\times\!   \frac{[3^\perp 4^\perp]^3}{[P_{12} 3^\perp ][4^\perp P_{12}]}
    \,+\,\frac{[2^\perp 3^\perp]^3}{[P_{23} 2^\perp][3^\perp P_{23}]}\!\times\!\frac{1}{P_{23}^2+m_{23}^2}\!\times\! \frac{[P_{23} 4^\perp]^3}{[4^\perp 1^\perp][1^\perp P_{23}]}\\[1ex]
    &=~
    \frac{\<q1^\perp\>^3}{\<q2^\perp\>\<q3^\perp\>\<q4^\perp\>}\biggl[\frac{[1^\perp 2^\perp][3^\perp 4^\perp]}{P_{12}^2+m_{12}^2}-
    \frac{[1^\perp 4^\perp][2^\perp 3^\perp]}{P_{23}^2+m_{23}^2}\biggr]\,,
\end{split}
\end{equation}
where
$m_{ij}^2=(m_i+m_j)^2$ is the mass$^2$  
 of the internal line.
Some comments are in order:
\begin{itemize}
  \item
  We
  have used little-group invariance on the internal line to convert all internal spinors to the CSW prescription,
  \begin{equation}\label{CSWpres}
    [\hat P_{I}^\perp|~\to~[ P_{I}|\equiv \<q|P_{I}\,.
  \end{equation}
  \item The shift parameters $c_i$ do not appear in the final answer, because the 3-point subamplitudes $\< X^- X^+\, X^+\>$ are
  anti-holomorphic
  in the spinors of external lines, and thus invariant under the holomorphic shift~(\ref{allline}).
  \item The amplitude $\< X^- X^+\, X^+\, X^+\>$ must of course vanish in
  the massless limit; indeed, when $m_i\to 0$ the two terms cancel against each other because momentum conservation in massless 4-point kinematics gives  $[34]/\<12\>-[14]/\<23\>=0$.
  \item $\< X^-\, X^+\, X^+\, X^+\>$  can be used to fix the normalization of the  4-point superamplitude with arbitrary massive lines; as we saw above the Grassmann structure is completely determined by supersymmetry already. This means that all other 4-point amplitudes are related to $\< X^-\, X^+\, X^+\, X^+\>$ by supersymmetry. We have verified numerically that the resulting 4-point superamplitude coincides with the result obtained from dimensional reduction
       and Grassmann Fourier transformation
      of the 6d superamplitude of~\cite{Bern:2010qa,Brandhuber:2010mm,Dennen:2010dh}.
\end{itemize}

In the all-line shift recursion relation for $\< X^-_1 X^+_2\,\cdots X_n^+\>$, all subamplitudes
are again
of the form $\< X^- X^+\,\cdots X^+\>$.
Note that $SU(2) \times SU(2)$ R-symmetry as well as our choice of all-$q$ equal do not admit any other subamplitudes to appear in the recursion diagrams.
Therefore, we can recursively use all-line shifts to reduce this amplitude into 3-point subamplitudes $\< X^- X^+X^+\>$.
Furthermore, since
$\< X^- X^+\, X^+\>=[2^\perp 3^\perp]^3/([1^\perp 2^\perp][3^\perp 1^\perp])$ is purely anti-holomorphic,
the
recursive reduction of $\< X^-X^+\,\cdots X^+\>$
fulfills all criteria stated in~\cite{Cohen:2010mi} for the validity of a massive CSW expansion --- it is an ``anti-MHV vertex expansion'' for $\< X^-_1 X^+_2\,\cdots X_n^+\>$:
\begin{equation}
    \< X^-_1 X^+_2\,\cdots X_n^+\>~=\sum_{{}^\text{\,\,\,\,\,\,cubic}_\text{diagrams}}
    \prod_{I} \frac{1}{P_I^2+m_I^2} \prod_{v} \frac{[v_2^\perp v_3^\perp]^3}{[v_1^\perp v_2^\perp][v_3^\perp v_1^\perp]}
   \,.
   \label{Acsw1}
\end{equation}
For
each diagram, the products run over the $n\!-\!2$ $\bMHV$ vertices $v$ and the $n\!-\!3$ internal lines $I$. The subamplitude of vertex $v$ is the three-point $\bMHV$ amplitude $\< X_{v_1}^- X_{v_2}^+X_{v_3}^+\>$, with the CSW prescription~(\ref{CSWpres}) understood for all internal-line spinors.

Let us now take this to the level of superamplitudes.
The amplitude \reef{Acsw1}
is sufficient to determine the full MHV superamplitude on the Coulomb-branch:
it is given by
\begin{equation}
    \A_n^{{\rm MHV} \times {\rm MHV}}~=~\frac{\< X^-_1 X^+_2\,\cdots X_n^+\>}{\<q1^\perp\>^4 K_2^2}\times \massA_{12}\times \massA_{34}\,,
\end{equation}
where $\massA_{12}$ was defined in~(\ref{massA12}), and the kinematic factors $K_i$ for arbitrary masses on the external lines were given in~(\ref{K246genM}).
The normalization follows simply from projecting out $\< X^-_1 X^+_2\,\cdots X_n^+\>$.

A crucial ingredient needed to derive the anti-MHV vertex expansion from the all-line shift recursion is the fact that $\bMHV\times\bMHV$ 3-point amplitudes $\< X^- X^+X^+\>$ are invariant under the shift~(\ref{allline}). But this invariance
extends beyond the $\bMHV$ level!
It is also true for 3-point
$ \overline{\text{MHV}}\times$MHV
amplitudes
as is obvious from their superamplitude~(\ref{SU4viol3pt}),
which
 contains angle spinors only in the invariant combination $\<qi^\perp\>$.
Thus any amplitude whose states only carry the $SU(2)$ indices $a=1,2$ once each, and arbitrary combinations of $a=3,4$ indices
is also computable from the CSW expansion.
Examples are amplitudes that contain one negative helicity gauge boson $X^-$ and an arbitrary number of scalars $x^{34}$ and positive helicity gauge bosons $X^+$.
Together with supersymmetry, this can  be used to determine the full MHV$\!\times\!$(anything) superamplitude. We will begin by stating the answer, then outline  its derivation.

The full MHV$\!\times\!$(anything) superamplitude, with arbitrarily many massive lines, takes the form
\begin{equation}\label{MHVanything}
\boxed{\phantom{\Biggl(}
    \A_n^\text{MHV$\!\times\!$(anything)}~=~\frac{\massA_{12}}{K_2}\times\!
    \sum_{{}^\text{\,\,\,\,\,\,cubic}_\text{diagrams}}
    \prod_{I} \int \!\!\frac{d\eta_{I3}\eta_{I4}}{P_I^2+m_I^2}
    \prod_{v} R_3
    \,.
    ~}
\end{equation}
In this superamplitude, the entire $\eta_{i1}, \eta_{i2}$ dependence is captured by $\massA_{12}$, while the $\eta_{i3}, \eta_{i4}$ dependence sits in the vertex superamplitudes $R_3$ 
given by
\begin{equation}\label{R3}
    R_3~=~
    \frac{\delta_{34}^{(2)}\big([1^\perp2^\perp]\eta_{3a}\!+\!\text{cyc}\big)}{\<q|1.2|q\>}
    ~+~
    \frac{m_2\<q|1|q]-m_1\<q|2|q]}{\<q|1|q]\<q|2|q]\<q|3|q]}
    \prod_{a=3}^4\!\!\Big([1^\perp q]\eta_{2a}\eta_{3a}+\text{cyc}\Big)\,.
\end{equation}
Each factor of $R_3$ in the product over vertices $v$ in~(\ref{MHVanything}) is evaluated with the momenta $1,2,3\to v_1,v_2,v_3$ corresponding to the vertex. As before, the CSW prescription~(\ref{CSWpres})
is implied for all square spinors that correspond to internal lines.

To derive~(\ref{MHVanything}), we proceed as follows.
The first step is to write the  MHV$\!\times\!$(anything) superamplitude in the  following form:
\begin{equation}\label{MHVanything2}
    \A_n^\text{MHV$\!\times\!$(anything)}=\frac{\massA_{12}}{K_2\<q1^\perp\>^2}\times
    \int d\eta_{11}d\eta_{12}\sum_{{}^\text{\,\,\,\,\,\,cubic}_\text{diagrams}}
    \prod_{I} \int \frac{d^4\eta_{Ia}}{P_I^2+m_I^2}
    \prod_{v} \bigl(\A^{\bMHV\!\times\!\bMHV}_3+\A^{\bMHV\!\times\!\text{MHV}}_3\bigr)\biggr|_{\eta_{i1},\eta_{i2}\to0}
    \,,
\end{equation}
where the $\A_3$ are the 3-point superamplitudes~\reef{3pointsuper} and (\ref{SU4viol3pt})
corresponding to vertex $v$, with the CSW prescription for spinors of internal lines understood.
Here, we denoted the $\bMHV$-band superamplitude pedantically by $\A^{\bMHV\!\times\!\bMHV}_3$  to avoid confusion.
To see that~(\ref{MHVanything2}) is indeed the correct superamplitude, note that the right part $\int d\eta_{11}d\eta_{12}[\ldots]|_{\eta_{i1},\eta_{i2}\to0}$ projects out amplitudes from the anti-MHV vertex expansion that carry R-symmetry indices $a=1,2$ only on line 1. As argued above, for all such amplitudes the anti-MHV vertex expansion is valid, irrespective of their structure with respect to the second $SU(2)$. This guarantees that the $\eta_{i3},\eta_{i4}$-structure of~(\ref{MHVanything}) is correct. The $\eta_{i1},\eta_{i2}$-structure is that of an MHV band, as the explicit
$\massA_{12}$
makes manifest. Finally, to check the overall normalization, we note that
\begin{equation}
    \int d\eta_{11}d\eta_{12}\frac{\massA_{12}}{K_2\<q1^\perp\>^2}\biggr|_{\eta_{i1},\eta_{i2}\to 0}\!\!\!\!=~1\,.
\end{equation}
This ensures that any amplitude  that carries R-symmetry indices $a=1,2$ only on line 1 is projected out correctly from $\A_n^\text{MHV$\!\times\!$(anything)}$, and thus confirms the overall normalization.

We now carry out the $\eta$-integration in the first $SU(2)$ sector. Note that 
\begin{equation}
    \A^{\bMHV}_3+\A^{\bMHV\!\times\!\text{MHV}}_3~=~\delta_{12}^{(2)}\big(\<qi^\perp\>\eta_{ia}\big)\times R_3\,,
\end{equation}
where $R_3$ is the superamplitude~(\ref{R3}) that only ``lives'' in the $SU(2)$ sector corresponding to $a=3,4$.
We 
carry out the integrations with respect to $\eta_{1a},\eta_{Ia}$ for $a=1,2$ in~(\ref{MHVanything2}),
giving a factor of $\<q1^\perp\>^2\prod_I\<q\hat P_I^\perp\>^2$.  The product $\prod_I\<q\hat P_i^\perp\>^2$ is canceled by the little-group scaling of the internal lines that allows us to replace the $|\hat P_i^\perp]$ of the internal line by its CSW prescription~(\ref{CSWpres}).
It follows that 
the full superamplitude
$\A_n^\text{MHV$\!\times\!$(anything)}$
takes the form
presented in~(\ref{MHVanything}).

\setcounter{equation}{0}
\section{Massive dual conformal invariance}
\label{s:dci}

The dual superconformal symmetry of planar amplitudes in $\cn=4$ SYM \cite{Drummond:2008vq,Brandhuber:2008pf} has enabled significant progress in understanding scattering amplitudes  at the origin of moduli space.  Although conventional conformal symmetry is broken on the Coulomb branch, it has been proposed \cite{Alday:2009zm, Henn:2010bk,Henn:2010ir,Henn:2011xk} that a form of {\it dual} conformal symmetry nonetheless persists. This is perhaps not completely surprising given the interesting recent observations that dual conformal symmetry is also a property of planar tree amplitudes in maximally SYM in 10d \cite{CaronHuot:2010rj} and 6d  \cite{Bern:2010qa,Brandhuber:2010mm,Dennen:2010dh}. For example, 6d superamplitudes
 at tree level 
transform covariantly under 6d dual conformal inversions, albeit with additional weight coming from the mismatch between the momentum and supermomentum delta functions in six dimensions \cite{Bern:2010qa}.
A complete understanding of dual conformal symmetry on the Coulomb branch of $\cn=4$ SYM could be a helpful guide for further progress on the structure of these amplitudes, at both tree and loop level.

Just as at the origin of moduli space, momentum conservation on the Coulomb branch suggests the definition of conventional region momenta,
\bea
\sum p_i = 0 ~~~\to~~~ (x_i - x_{i+1})_{\alpha \dot \alpha} \,\equiv\, (x_{i,i+1})_{\alpha \dot \alpha}\, = \,|i^\perp\> [i^\perp| + \mu_i \bar \mu_i |q\> [q|\,,
\eea
with the periodic identification $x_{n+1} = x_1$.
Likewise, the Coulomb branch condition $\sum m_i = 0$ (from the 6d perspective, conservation of momentum in the extra dimensions) suggests `region masses',\footnote{Here, as elsewhere, we are taking $m_i$ real. Properly speaking,
 $n_{i,i+1} = Z_i,$ 
but the reality condition allows us to identify the region masses directly with $m_i$.}
\bea
\sum m_i = 0 ~~\to~~ (n_i - n_{i+1}) \equiv n_{i,i+1} = m_i\,.
\eea
These region variables may be collected together into a larger object  suggestive of the higher-dimensional origin: let us define $\hat x_i = (x_i, n_i)$ such that $\hat x_{i,i+1}^2 = x_{i,i+1}^2 + n_{i,i+1}^2$.

We are going to explore the behavior of certain planar Coulomb branch amplitudes under dual conformal inversions. The inversions act on $x_i$ and $m_i$ simply
 as~\cite{Alday:2009zm} 
\bea 
I\bigl[x_{i,i+1}^2\bigr] \,=\,\frac{x_{i,i+1}^2}{\hat x_i^2 \hat x_{i+1}^2}\,,~~ \qquad I\bigl[m_i^2\bigr] \,=\,\frac{m_i^2}{\hat x_i^2 \hat x_{i+1}^2}\,.
\eea
Less trivially, the inversion properties of helicity spinors for massive momenta may be obtained by carrying out the reduction of the corresponding six-dimensional expressions found in \cite{Bern:2010qa,Brandhuber:2010mm,Dennen:2010dh}.\footnote{It is important to note that in order to reproduce the known 4d inversion properties in the massless limit, the 6d inversions should not raise or lower the 6d little group index.
} For example, we have (spinor indices implicit):
\bea
I\bigl[\<i^\perp|\bigr] \,=\, \frac{1}{\hat x_i^2} \left( x_i |i^\perp\> - n_i \mu_i |q]\right)\,,~~
 \qquad I\bigl[[i^\perp|\bigr] \,=\,\frac{1}{\hat x_i^2} \left( x_i |i^\perp] - n_i \bar \mu_i |q \> \right)\,,
 \\ \nonumber
I\bigl[\bar \mu_i \< q|\bigr] \,=\,\frac{1}{\hat x_i^2} \left( \bar \mu_i x_i |q \> + n_i |i^\perp] \right)\,,~~
  \qquad I\bigl[ \mu_i [q|\bigr] \,=\,\frac{1}{\hat x_i^2} \left( \mu_i x_i |q] + n_i |i^\perp \> \right)\,.
\eea

Let us now apply dual inversion  to the $w^{\perp}\overline{w}^\perp$-gluon UHV amplitude
 from the MHV-band, where $w^\perp$ denotes the scalar $w^\perp=(w^{12}-w^{34})/\sqrt{2}$ ``orthogonal'' to the longitudinal gauge boson.
 As both massive lines in this amplitude are scalars, the resulting amplitude must be independent of the reference spinor $q$. Indeed,
 projecting out $w^{\perp}$ and $\overline w^{\perp}$ on lines 1 and 2 from the superamplitude \reef{AMHV} and positive helicity gluons on the other lines, we find
\bea
\label{wwggamp}
\big\< w^{\perp}_1 \, \overline w^{\perp}_2 \, g_3^+ \, g_4^+ \dots g_n^+ \big\> ~=~
\frac{
 m^2 \,
[3|\prod_{i=4}^{n-1}[m^2-x_{i2}x_{2,i+1}]|n]}{\<34\>\<45\>\cdots \<n\!-\!1,n\>\prod_{i=4}^{n}(x_{2 i}^2+m^2)}
   \,.
\eea
The dual conformal properties of \reef{wwggamp} are not obvious. However, in this case with just two massive lines, the region masses are such that $m = n_{2i}$ for any $i=1,3,4,\dots, n$. Hence we can rewrite the propagators using $x_{2i}^2 + m^2 = x_{2i}^2 + n_{2i}^2 = \hat x_{2i}^2$. In the numerators we write
 $(m^2 - x_{i2} x_{2,i+1})_{\alpha}^{~\,\beta} = (-\hat x_{i2} \hat x_{2,i+1})_{\alpha}^{~\,\beta}$,
 which is the projection of a 6d spinor product to 4d square spinors.
 Concretely, writing $\hat x_{ij}$ in terms of 4d Dirac matrices as $\hat x_{ij}=x_{ij}^\mu \gamma_\mu+n_{ij}\gamma_5$, this projection can be implemented as a chiral projection $P_-=(1-\gamma_5)/2$.
This now allows us to write the amplitude \reef{wwggamp} as
\begin{equation}
\boxed{
\begin{split} 
\big\< w^{\perp}_1 \, \overline w^{\perp}_2  \, g_3^+ \, g_4^+ \dots g_n^+ \big\>
   ~&=~
   \frac{
   (-)^{n}\, 
   m^2\,[3|
   \hat x_{32} \hat x_{25}
   \sep
   \hat x_{52} \hat x_{26}
   \sep\cdots\sep\hat x_{n\dash1,2}\hat x_{2n}| n]}{\<34\>\<45\>\cdots \<n\!-\!1,n\>\prod_{i=4}^{n}\hat x_{2i}^2}\,,
\end{split}
}
\label{wwggamp2}
\end{equation}
where the separators $\sep$ indicate
 the projection $P_-$ onto square spinors 
(and thus prevents us from extracting propagators $\hat x_{2 i}^2$ from the string of spinors).

Under dual conformal inversion, the string $\< 34 \> \<45 \> \cdots \<n-1, n\>$ is covariant with weight $(\hat x_3^2 \cdots \hat x_{n-1}^2)^{-1}$, just as in the massless case. The propagators $\prod_{i=4}^n \hat x_{2i}^2$  transform covariantly with weight $(\hat x_2^{2n -6} \hat x_4^2 \cdots \hat x_n^2)^{-1}$, while the spinor string $[3|\hat x_{32}\cdots\hat x_{2n}|n]$ transforms covariantly with weight $(\hat x_2^{2n-8} \hat x_3^2 x_5^2 \cdots \hat x_{n}^2)^{-1}$. Note that there is an ambiguity as to which weight to assign to the $m^2$-factor. Let us for now  arbitrarily assign $m^2 = n_{12}^2$, so that it transforms with weight $(\hat x_1^2 \hat x_2^2)^{-1}$. Finally we recall that the momentum delta-function has weight 4 at $\hat{x}_1$. Thus, all in all, the amplitude \reef{wwggamp2} transforms covariantly under dual conformal inversions, but with abnormal weight
$
\hat{x}_1^6 \hat{x}_4^4 \hat{x}_5^2\cdots \hat x_{n-1}^2
=(\hat{x}_1^2 \hat{x}_2^2 \cdots \hat{x}_n^2)
\frac{\hat{x}_1^4 \hat{x}_4^2}{\hat{x}_2^2 \hat{x}_3^2 \hat{x}_n^2}$.
The unusual weights reflect our simplifying choices for amplitudes with just two massive lines and the resulting ambiguity in the mass region variables;
the overall weights can the changed using $1 = m^2/m^2 = n_{2i}^2/n_{2j}^2$. The main point here is not to understand these weights, but to illustrate that there is a way to write the amplitude
 in a way that is {\em manifestly dual conformal covariant, albeit with ambiguous weights}.

In the case of massless planar amplitudes, individual amplitudes are not generically dual conformal invariant; only split-helicity amplitudes transform covariantly, and in general one needs to promote the amplitudes to superamplitudes for covariance to become manifest. The  amplitude \reef{wwggamp2} is in a sense a massive  analogue of the split-helicity amplitudes at the origin of moduli space. If we try to press this analogy further, we encounter a new complication. The
amplitude $\<W^- \overline{W}^+ g^+\dots g^+\>$ is closely related to the
 $w^{\perp} \overline w^{\perp}$-gluon 
amplitude \reef{wwggamp2}; they differ only by an overall factor
 $-\<q 1^\perp \>^2/\<q 2^\perp \>^2$  
which carries the helicity weights of the vectors. This factor, however, does \emph{not} transform covariantly under dual conformal inversions for general $q$. Curiously, though, our special choice  $q=q_0$,
in \reef{s:smart} implies that\footnote{ 
To see this, note that $\<qi^\perp\>=[i^\perp q]$, a relation that we did not make use of in this paper so far, but that was discussed extensively in~\cite{Cohen:2010mi}.}
 $\<q_0 1^\perp \>^2=-\<q_0 2^\perp \>^2$ 
so that the non-covariant factor is eliminated and the $W\overline{W}$-gluon amplitude exactly coincides with the
 $w^{\perp} \overline w^{\perp}$-gluon 
amplitude \reef{wwggamp2}. In other words, the special choice of $q_0$ erases the information from the transverse polarization vectors. The point to beware of here is, however, that the dual conformal transformations do not generally leave
 invariant the
 $q$-basis 
 chosen for the helicity amplitudes. 
In the massless case this is not an issue because the massless helicity amplitudes are frame-independent. But for the massive amplitudes it is crucial, and any generic choice of $q$'s cannot be expected to be preserved, or preferred, by the dual conformal symmetry. Thus the massive amplitudes that have a chance of displaying manifest covariance under dual conformal inversions are those which do not depend explicitly on $q$, such as for example \reef{wwggamp2}.\footnote{
This is consistent with the evidence in~\cite{Alday:2009zm} that Coulomb-branch amplitudes are invariant under dual conformal transformations when not taking into account contributions from the polarization of external states.}

It is natural to suspect that the dual conformal properties of the Coulomb branch are more transparent at the level of superamplitudes. Supermomentum conservation allows us to define Grassmann region variables such as
\bea
|\theta_{1i}\> - |\theta_{1i+1}\> \equiv |\theta_{1 i,i+1}\>= |i^\perp \> \eta_{1i} - \bar \mu_i |q\>
 \frac{\partial}{\partial \eta_{i2}} \, ,
\eea
and likewise for the remaining supercharges in each of the two $SU(2)$ sectors. Under inversions these region supermomenta transform as, e.g.,
\bea
I\big[|\theta_{1i}\>\big] \,=\, \frac{1}{\hat x_i^2} \left( \< \theta_{1i} | x_i - n_i [\theta^2_i| \right) \, .
\eea
The Grassmann variables $\eta$ transform inhomogeneously under dual conformal inversion. It is therefore not clear that each of our N$^k$MHV-bands will have dual superconformal symmetry, or whether it will only be a property of the full $n$-point superamplitudes. Let us note, however, that the 4-point MHV-band superamplitude (\ref{AMHV}) is expected to transform covariantly. This follows from 6d dual conformal invariance since $\massA_{12} \times \massA_{34}$ for $n=4$ is related to the 6d supermomentum delta functions
 by 
a
Grassmann Fourier transform [\citen{Hatsuda:2008pm}, \citen{Cohen:2010mi}], generalized to the massive case.
The simple connection between the
 4d SUSY invariant $\massA_{12} \times \massA_{34}$  and the 6d invariant $\delta^8(Q)$ 
only holds for four external lines. Thus the explicit behavior of massive N$^k$MHV superamplitude-bands with $n \geq 5$ under dual conformal transformations remains an open question.

%%%%%%%%%%%%%%

It is interesting to consider the infinitesimal massive dual conformal boost generator acting on the full set of Coulomb branch coordinates. Decomposing the corresponding 6d generators, we find that the massive dual conformal boost generators take the form (spinor indices implicit)
\bea \nonumber
K_{\dot \alpha}^{\; \alpha} &=& \sum_{i=1}^n \left \{ x_i x_i \frac{\partial }{\partial x_i} + 2 x_i n_i \frac{\partial}{\partial n_i} + n_i^2 \frac{\partial}{\partial x_i} \right. \\
&&\qquad~+ \bar \mu_i \< q| x_i \frac{\partial}{\partial(\bar \mu_i |q\>)} + \<i^\perp| x_i \frac{\partial}{\partial |i^\perp \>} +  \mu_i [q| x_i  \frac{\partial}{\partial( \mu_i |q])} + [i^\perp| x_i \frac{\partial}{\partial |i^\perp ]}
 \label{Kgen}
\\
\nonumber
 &&\qquad~+ \left. n_i \bar \mu_i \<q| \frac{\partial}{\partial |i^\perp]} - n_i \< i^\perp| \frac{\partial}{\partial(\mu_i |q])} + n_i [i^\perp| \frac{\partial}{\partial (\bar \mu_i |q \>)} -  n_i  \mu_i \frac{\partial}{\partial |i^\perp \>} \right \}\,.
\eea
The first line
has the familiar bosonic components appearing in \cite{Alday:2009zm, Bern:2010qa}. The remaining two lines give the transformation properties of the helicity spinors.  The massless limit reduces precisely to the dual conformal boost generator at the origin of moduli space \cite{Drummond:2008vq,Brandhuber:2008pf}. In essence, the action of the massive dual conformal boost generator on amplitudes provides a differential equation relating amplitudes at different points on moduli space.
It would be exciting to realize a useful set of differential equations to move amplitudes around on the moduli-space.

Let us finally note that in order to fully understand dual conformal symmetry on the Coulomb branch, it may be useful to relax the condition $q_i=q$ which simplified our results significantly in the earlier sections.
This allows a more general framework with generic reference vectors, and working also with generic masses $m_i$ subject to $\sum_{i=1}^n m_i =0$ may lead to a more natural implementation of dual conformal symmetry. Our approach here was exploratory in the context of the formalism used in the previous sections, and we hope
for further progress  in the future.

%%%%%%%%%%%%%%
%%%%%%%%%%%%%%
%%%%%%%%%%%%%%
\setcounter{equation}{0}
\section{Conclusion and Outlook}
\label{s:con}
In this paper, we presented a wide variety of compact expressions for massive amplitudes and superamplitudes, as well as a simple prescription for generating Coulomb-branch amplitudes from the origin of moduli space. We have also shown that some amplitudes can be compactly written in a manifestly dual conformal form.
Taken together, our results suggest that much of the simplicity apparent at the origin of moduli space persists as we move onto the Coulomb branch.

Our analysis in this paper is far from exhaustive and should be regarded as a first exploratory step onto the Coulomb branch. For example, using the soft-limit method introduced here, it seems within reach to obtain compact and simple expressions for the entire Coulomb-branch superamplitude to leading order in the mass. The soft-limit method may also be applicable beyond leading order; however, in this case, new subtleties arise that must be addressed. In particular, the soft-limits that
are naturally associated with sub-leading mass corrections to Coulomb-branch amplitudes suffer from
soft divergences that are absent in our leading-order analysis.
As an example, consider $\< W^+\,  \,\overline W^+\, g^-\, g^-\>$.
At leading order in $m\to 0$, we recover the usual massless Parke-Taylor amplitude. One might hope that the subleading terms of $O(m^2)$ are captured by the double-soft limit of $\< g^+\, \phi^{12}_{q}\, \phi^{34}_{q} \,g^+\, g^-\, g^-\>$. Symmetrization of the scalars eliminates the collinear divergence, as in section \ref{s:soft}, but a soft divergence proportional to $1/s_{1q}+1/s_{2q}$ remains as $q\to 0$.
This leads to ambiguous answers for the finite remaining term that we
wish to extract. Curiously, the ``special choice of $q$'' that we introduced
in section \ref{s:smart}
to simplify the
superamplitudes may  play a role in extracting the
correct
subleading behavior. In fact, in
the above
example
the soft divergence
vanishes for $q=q_0$. If one can systematically obtain the subleading terms in the small-mass expansion of tree-amplitudes, then unitarity methods allow one to extract the subleading mass-corrections also at loop-level. This could be useful in phenomenological applications where the energy scale of the scattering process dominates the masses.

It has been observed in planar massless $\cn=4$ SYM that, at the level of the integrand, increasing the loop level is very closely related to an increase in N$^k$MHV level~\cite{CaronHuot:2010ek,Mason:2010yk}. If our soft-limit prescription can be extended beyond the leading order, it will exhibit a similar ``conservation of complexity'': higher-order mass corrections are associated with soft-limits of massless amplitudes of higher N$^k$MHV level --- again there is no free lunch.
This fits well with the suggestion of~\cite{Elvang:2011fx} that 6d SYM amplitudes at the $n$-point level may be reconstructible from the full {\em massless} 4d superamplitude, including the alternating-helicity sector amplitudes, which are the most complicated ones for the given $n$. The 6d amplitudes in turn determine the massive Coulomb-branch amplitudes completely.

There is a third avenue onto the Coulomb-branch: imposing that amplitudes are annihilated by dual conformal generators such as~(\ref{Kgen})
implies a differential equation that connects different points in moduli space. In particular, one should in principle be able to determine the complete mass dependence from the leading one. It would be valuable to elucidate the connection between these three seemingly very different approaches to obtain Coulomb-branch amplitudes from massless ones.

The discussion in this paper has focused entirely on tree amplitudes. Coulomb-branch amplitudes
with massless external states only
have
been studied at
loop-level
to regulate integrals of the massless theory without spoiling dual conformal properties. It would be useful to study loop-level (super)amplitudes on the Coulomb branch in their own right to see how many of the nice results obtained at the origin of moduli space can persist in a massive non-conformal theory. As the general  superamplitude structure discussed in this paper (such as the SUSY invariant $\massA_{12}\times \massA_{34}$) is valid at both
tree- and loop-level,
the results presented here should facilitate an extension to loop amplitudes.

It would be useful to apply the techniques developed here toward the computation of QCD amplitudes involving massive quarks.
Much as massless QCD amplitudes may be readily extracted from $\cn = 4$ superamplitudes at the origin of moduli space \cite{Dixon:2010ik}, so too may massive QCD amplitudes be extracted from our Coulomb-branch superamplitudes. 
This could be very useful for
efficient computation
of processes relevant for collider physics.

\vskip .1in

\noindent  {\bf Acknowledgements}

We are grateful to Nima Arkani-Hamed for early collaboration and for numerous helpful discussions. We also thank Simon Caron-Huot, Yu-tin Huang and Michael Peskin for helpful discussions.
We thank Rutger Boels and Christian Schwinn for discussions and coordination of their related work~\cite{BoelsSchwinn}.
NC is
supported in part by the NSF under grant PHY-0907744.
HE is supported by NSF CAREER Grant PHY-0953232, and in part by the US Department of Energy under DOE grants DE-FG02-95ER 40899. During the early stages of this project, HE was also supported by the Institute for Advanced Study (DOE grant DE-FG02-90ER40542).
The research of MK is supported by NSF grant PHY-0756966. TS is supported in part by the US Department of Energy under DOE grant DE-FG02-90ER40542 and the NSF under grant AST-0807444. NC and TS gratefully acknowledge support from the Institute for Advanced Study.

\renewcommand{\baselinestretch}{1.0}
\small
\setlength{\parskip}{0 pt}

%\bibliographystyle{utphys}
%\bibliography{coulombbib}{}

\begin{thebibliography}{10}
\bibitem{Drummond:2008vq}
J.~Drummond, J.~Henn, G.~Korchemsky, and E.~Sokatchev, ``{Dual superconformal
  symmetry of scattering amplitudes in $N=4$ super-Yang-Mills theory},''
  \href{http://dx.doi.org/10.1016/j.nuclphysb.2009.11.022}{{\em Nucl.Phys.}
  {\bf B828} (2010)  317--374}, \href{http://arxiv.org/abs/0807.1095}{{\tt
  arXiv:0807.1095 [hep-th]}}.

\bibitem{Brandhuber:2008pf}
A.~Brandhuber, P.~Heslop, and G.~Travaglini, ``{A Note on dual superconformal
  symmetry of the N=4 super Yang-Mills S-matrix},''
  \href{http://dx.doi.org/10.1103/PhysRevD.78.125005}{{\em Phys.Rev.} {\bf D78}
  (2008)  125005}, \href{http://arxiv.org/abs/0807.4097}{{\tt arXiv:0807.4097
  [hep-th]}}.

\bibitem{Drummond:2008bq}
J.~Drummond, J.~Henn, G.~Korchemsky, and E.~Sokatchev, ``{Generalized unitarity
  for $N=4$ super-amplitudes},'' \href{http://arxiv.org/abs/0808.0491}{{\tt
  arXiv:0808.0491 [hep-th]}}.

\bibitem{Drummond:2008cr}
J.~M. Drummond and J.~M. Henn, ``{All tree-level amplitudes in $N=4$ SYM},''
  \href{http://dx.doi.org/10.1088/1126-6708/2009/04/018}{{\em JHEP} {\bf 04}
  (2009)  018},
\href{http://arxiv.org/abs/0808.2475}{{\tt arXiv:0808.2475 [hep-th]}}.
%%CITATION = 0808.2475;%%.

\bibitem{Brandhuber:2009xz}
A.~Brandhuber, P.~Heslop, and G.~Travaglini, ``{One-Loop Amplitudes in $N=4$
  Super Yang-Mills and Anomalous Dual Conformal Symmetry},''
  \href{http://dx.doi.org/10.1088/1126-6708/2009/08/095}{{\em JHEP} {\bf 0908}
  (2009)  095}, \href{http://arxiv.org/abs/0905.4377}{{\tt arXiv:0905.4377
  [hep-th]}}.

\bibitem{Elvang:2009ya}
H.~Elvang, D.~Z. Freedman, and M.~Kiermaier, ``{Dual conformal symmetry of
  1-loop NMHV amplitudes in $N=4$ SYM theory},''
  \href{http://dx.doi.org/10.1007/JHEP03(2010)075}{{\em JHEP} {\bf 1003} (2010)
   075}, \href{http://arxiv.org/abs/0905.4379}{{\tt arXiv:0905.4379 [hep-th]}}.

\bibitem{ArkaniHamed:2009vw}
N.~Arkani-Hamed, F.~Cachazo, and C.~Cheung, ``{The Grassmannian Origin Of Dual
  Superconformal Invariance},''
  \href{http://dx.doi.org/10.1007/JHEP03(2010)036}{{\em JHEP} {\bf 1003} (2010)
   036}, \href{http://arxiv.org/abs/0909.0483}{{\tt arXiv:0909.0483 [hep-th]}}.

\bibitem{ArkaniHamed:2010kv}
N.~Arkani-Hamed, J.~L. Bourjaily, F.~Cachazo, S.~Caron-Huot, and J.~Trnka,
  ``{The All-Loop Integrand For Scattering Amplitudes in Planar $N=4$ SYM},''
  \href{http://dx.doi.org/10.1007/JHEP01(2011)041}{{\em JHEP} {\bf 1101} (2011)
   041}, \href{http://arxiv.org/abs/1008.2958}{{\tt arXiv:1008.2958 [hep-th]}}.

\bibitem{Mason:2010yk}
L.~Mason and D.~Skinner, ``{The Complete Planar S-matrix of $N=4$ SYM as a
  Wilson Loop in Twistor Space},''
  \href{http://dx.doi.org/10.1007/JHEP12(2010)018}{{\em JHEP} {\bf 1012} (2010)
   018}, \href{http://arxiv.org/abs/1009.2225}{{\tt arXiv:1009.2225 [hep-th]}}.

\bibitem{Goncharov:2010jf}
A.~B. Goncharov, M.~Spradlin, C.~Vergu, and A.~Volovich, ``{Classical
  Polylogarithms for Amplitudes and Wilson Loops},''
  \href{http://dx.doi.org/10.1103/PhysRevLett.105.151605}{{\em Phys.Rev.Lett.}
  {\bf 105} (2010)  151605}, \href{http://arxiv.org/abs/1006.5703}{{\tt
  arXiv:1006.5703 [hep-th]}}.

\bibitem{Gaiotto:2011dt}
D.~Gaiotto, J.~Maldacena, A.~Sever, and P.~Vieira, ``{Pulling the straps of
  polygons},'' \href{http://arxiv.org/abs/1102.0062}{{\tt arXiv:1102.0062
  [hep-th]}}.

\bibitem{Alday:2009zm}
L.~F. Alday, J.~M. Henn, J.~Plefka, and T.~Schuster, ``{Scattering into the
  fifth dimension of $N=4$ super Yang-Mills},''
  \href{http://dx.doi.org/10.1007/JHEP01(2010)077}{{\em JHEP} {\bf 1001} (2010)
   077}, \href{http://arxiv.org/abs/0908.0684}{{\tt arXiv:0908.0684 [hep-th]}}.

\bibitem{Henn:2010bk}
J.~M. Henn, S.~G. Naculich, H.~J. Schnitzer, and M.~Spradlin,
  ``{Higgs-regularized three-loop four-gluon amplitude in $N=4$ SYM:
  exponentiation and Regge limits},''
  \href{http://dx.doi.org/10.1007/JHEP04(2010)038}{{\em JHEP} {\bf 1004} (2010)
   038}, \href{http://arxiv.org/abs/1001.1358}{{\tt arXiv:1001.1358 [hep-th]}}.

\bibitem{Henn:2010ir}
J.~M. Henn, S.~G. Naculich, H.~J. Schnitzer, and M.~Spradlin, ``{More loops and
  legs in Higgs-regulated $N=4$ SYM amplitudes},''
  \href{http://dx.doi.org/10.1007/JHEP08(2010)002}{{\em JHEP} {\bf 1008} (2010)
   002}, \href{http://arxiv.org/abs/1004.5381}{{\tt arXiv:1004.5381 [hep-th]}}.

\bibitem{Henn:2011xk}
J.~M. Henn, ``{Dual conformal symmetry at loop level: massive
  regularization},'' \href{http://arxiv.org/abs/1103.1016}{{\tt arXiv:1103.1016
  [hep-th]}}.

\bibitem{Schabinger:2008ah}
R.~M. Schabinger, ``{Scattering on the Moduli Space of $N=4$ Super
  Yang-Mills},'' \href{http://arxiv.org/abs/0801.1542}{{\tt arXiv:0801.1542
  [hep-th]}}.

\bibitem{Boels:2010mj}
R.~H. Boels, ``{No triangles on the moduli space of maximally supersymmetric
  gauge theory},'' \href{http://dx.doi.org/10.1007/JHEP05(2010)046}{{\em JHEP}
  {\bf 05} (2010)  046},
\href{http://arxiv.org/abs/1003.2989}{{\tt arXiv:1003.2989 [hep-th]}}.
%%CITATION = 1003.2989;%%.

\bibitem{Dixon:2004za}
L.~J. Dixon, E.~W.~N. Glover, and V.~V. Khoze, ``{MHV rules for Higgs plus
  multi-gluon amplitudes},''
  \href{http://dx.doi.org/10.1088/1126-6708/2004/12/015}{{\em JHEP} {\bf 12}
  (2004)  015},
\href{http://arxiv.org/abs/hep-th/0411092}{{\tt arXiv:hep-th/0411092}}.
%%CITATION = HEP-TH/0411092;%%.

\bibitem{Badger:2004ty}
S.~D. Badger, E.~W.~N. Glover, and V.~V. Khoze, ``{MHV rules for Higgs plus
  multi-parton amplitudes},''
  \href{http://dx.doi.org/10.1088/1126-6708/2005/03/023}{{\em JHEP} {\bf 03}
  (2005)  023},
\href{http://arxiv.org/abs/hep-th/0412275}{{\tt arXiv:hep-th/0412275}}.
%%CITATION = HEP-TH/0412275;%%.

\bibitem{Berger:2006sh}
C.~F. Berger, V.~Del~Duca, and L.~J. Dixon, ``{Recursive construction of
  Higgs+multiparton loop amplitudes: The last of the phi-nite loop
  amplitudes},'' \href{http://dx.doi.org/10.1103/PhysRevD.74.094021}{{\em Phys.
  Rev.} {\bf D74} (2006)  094021},
\href{http://arxiv.org/abs/hep-ph/0608180}{{\tt arXiv:hep-ph/0608180}}.
%%CITATION = HEP-PH/0608180;%%.

\bibitem{Badger:2007si}
S.~D. Badger, E.~W.~N. Glover, and K.~Risager, ``{One-loop phi-MHV amplitudes
  using the unitarity bootstrap},''
  \href{http://dx.doi.org/10.1088/1126-6708/2007/07/066}{{\em JHEP} {\bf 07}
  (2007)  066},
\href{http://arxiv.org/abs/0704.3914}{{\tt arXiv:0704.3914 [hep-ph]}}.
%%CITATION = 0704.3914;%%.

\bibitem{Dixon:2009uk}
L.~J. Dixon and Y.~Sofianatos, ``{Analytic one-loop amplitudes for a Higgs
  boson plus four partons},''
  \href{http://dx.doi.org/10.1088/1126-6708/2009/08/058}{{\em JHEP} {\bf 08}
  (2009)  058},
\href{http://arxiv.org/abs/0906.0008}{{\tt arXiv:0906.0008 [hep-ph]}}.
%%CITATION = 0906.0008;%%.

\bibitem{Badger:2009hw}
S.~Badger, E.~W. Nigel~Glover, P.~Mastrolia, and C.~Williams, ``{One-loop Higgs
  plus four gluon amplitudes: Full analytic results},''
  \href{http://dx.doi.org/10.1007/JHEP01(2010)036}{{\em JHEP} {\bf 01} (2010)
  036},
\href{http://arxiv.org/abs/0909.4475}{{\tt arXiv:0909.4475 [hep-ph]}}.
%%CITATION = 0909.4475;%%.

\bibitem{Boels:2008du}
R.~Boels and C.~Schwinn, ``{CSW rules for massive matter legs and glue
  loops},'' \href{http://dx.doi.org/10.1016/j.nuclphysbps.2008.09.094}{{\em
  Nucl. Phys. Proc. Suppl.} {\bf 183} (2008)  137--142},
\href{http://arxiv.org/abs/0805.4577}{{\tt arXiv:0805.4577 [hep-th]}}.
%%CITATION = 0805.4577;%%.

\bibitem{Boels:2009bv}
R.~Boels, ``{Covariant representation theory of the Poincare algebra and some
  of its extensions},'' \href{http://dx.doi.org/10.1007/JHEP01(2010)010}{{\em
  JHEP} {\bf 01} (2010)  010},
\href{http://arxiv.org/abs/0908.0738}{{\tt arXiv:0908.0738 [hep-th]}}.
%%CITATION = 0908.0738;%%.

\bibitem{Badger:2005zh}
S.~D. Badger, E.~W.~N. Glover, V.~V. Khoze, and P.~Svrcek, ``{Recursion
  Relations for Gauge Theory Amplitudes with Massive Particles},''
  \href{http://dx.doi.org/10.1088/1126-6708/2005/07/025}{{\em JHEP} {\bf 07}
  (2005)  025},
\href{http://arxiv.org/abs/hep-th/0504159}{{\tt arXiv:hep-th/0504159}}.
%%CITATION = HEP-TH/0504159;%%.

\bibitem{Schwinn:2007ee}
C.~Schwinn and S.~Weinzierl, ``{On-shell recursion relations for all Born QCD
  amplitudes},'' \href{http://dx.doi.org/10.1088/1126-6708/2007/04/072}{{\em
  JHEP} {\bf 0704} (2007)  072},
  \href{http://arxiv.org/abs/hep-ph/0703021}{{\tt arXiv:hep-ph/0703021
  [HEP-PH]}}.

\bibitem{Britto:2004ap}
R.~Britto, F.~Cachazo, and B.~Feng, ``{New Recursion Relations for Tree
  Amplitudes of Gluons},''
  \href{http://dx.doi.org/10.1016/j.nuclphysb.2005.02.030}{{\em Nucl. Phys.}
  {\bf B715} (2005)  499--522},
\href{http://arxiv.org/abs/hep-th/0412308}{{\tt arXiv:hep-th/0412308}}.
%%CITATION = HEP-TH/0412308;%%.

\bibitem{Britto:2005fq}
R.~Britto, F.~Cachazo, B.~Feng, and E.~Witten, ``{Direct Proof Of Tree-Level
  Recursion Relation In Yang- Mills Theory},''
  \href{http://dx.doi.org/10.1103/PhysRevLett.94.181602}{{\em Phys. Rev. Lett.}
  {\bf 94} (2005)  181602},
\href{http://arxiv.org/abs/hep-th/0501052}{{\tt arXiv:hep-th/0501052}}.
%%CITATION = HEP-TH/0501052;%%.

\bibitem{Forde:2005ue}
D.~Forde and D.~A. Kosower, ``{All-multiplicity amplitudes with massive
  scalars},'' \href{http://dx.doi.org/10.1103/PhysRevD.73.065007}{{\em Phys.
  Rev.} {\bf D73} (2006)  065007},
\href{http://arxiv.org/abs/hep-th/0507292}{{\tt arXiv:hep-th/0507292}}.
%%CITATION = HEP-TH/0507292;%%.

\bibitem{Rodrigo:2005eu}
G.~Rodrigo, ``{Multigluonic scattering amplitudes of heavy quarks},''
  \href{http://dx.doi.org/10.1088/1126-6708/2005/09/079}{{\em JHEP} {\bf 0509}
  (2005)  079}, \href{http://arxiv.org/abs/hep-ph/0508138}{{\tt
  arXiv:hep-ph/0508138 [hep-ph]}}.

\bibitem{Dittmaier:1998nn}
S.~Dittmaier, ``{Weyl-van-der-Waerden formalism for helicity amplitudes of
  massive particles},''
  \href{http://dx.doi.org/10.1103/PhysRevD.59.016007}{{\em Phys. Rev.} {\bf
  D59} (1999)  016007},
\href{http://arxiv.org/abs/hep-ph/9805445}{{\tt arXiv:hep-ph/9805445}}.
%%CITATION = HEP-PH/9805445;%%.

\bibitem{Cohen:2010mi}
T.~Cohen, H.~Elvang, and M.~Kiermaier, ``{On-shell constructibility of tree
  amplitudes in general field theories},''
  \href{http://arxiv.org/abs/1010.0257}{{\tt arXiv:1010.0257 [hep-th]}}.

\bibitem{YutinHuang}
Y.-t. Huang, ``{Non-Chiral S-Matrix of N = 4 Super Yang-Mills},'' {\em to
  appear}  .

\bibitem{Bern:2010qa}
Z.~Bern, J.~J. Carrasco, T.~Dennen, Y.-t. Huang, and H.~Ita, ``{Generalized
  Unitarity and Six-Dimensional Helicity},''
  \href{http://arxiv.org/abs/1010.0494}{{\tt arXiv:1010.0494 [hep-th]}}.

\bibitem{Brandhuber:2010mm}
A.~Brandhuber, D.~Korres, D.~Koschade, and G.~Travaglini, ``{One-loop
  Amplitudes in Six-Dimensional (1,1) Theories from Generalised Unitarity},''
  \href{http://dx.doi.org/10.1007/JHEP02(2011)077}{{\em JHEP} {\bf 02} (2011)
  077},
\href{http://arxiv.org/abs/1010.1515}{{\tt arXiv:1010.1515 [hep-th]}}.
%%CITATION = 1010.1515;%%.

\bibitem{Dennen:2010dh}
T.~Dennen and Y.-t. Huang, ``{Dual Conformal Properties of Six-Dimensional
  Maximal Super Yang-Mills Amplitudes},''
  \href{http://dx.doi.org/10.1007/JHEP01(2011)140}{{\em JHEP} {\bf 1101} (2011)
   140}, \href{http://arxiv.org/abs/1010.5874}{{\tt arXiv:1010.5874 [hep-th]}}.

\bibitem{Schwinn:2006ca}
C.~Schwinn and S.~Weinzierl, ``{SUSY Ward identities for multi-gluon helicity
  amplitudes with massive quarks},''
  \href{http://dx.doi.org/10.1088/1126-6708/2006/03/030}{{\em JHEP} {\bf 03}
  (2006)  030},
\href{http://arxiv.org/abs/hep-th/0602012}{{\tt arXiv:hep-th/0602012}}.
%%CITATION = HEP-TH/0602012;%%.

\bibitem{Parke:1986gb}
S.~J. Parke and T.~Taylor, ``{An Amplitude for $n$ Gluon Scattering},''
  \href{http://dx.doi.org/10.1103/PhysRevLett.56.2459}{{\em Phys.Rev.Lett.}
  {\bf 56} (1986)  2459}.

\bibitem{Witten:1978mh}
E.~Witten and D.~I. Olive, ``{Supersymmetry Algebras That Include Topological
  Charges},'' \href{http://dx.doi.org/10.1016/0370-2693(78)90357-X}{{\em
  Phys.Lett.} {\bf B78} (1978)  97}.

\bibitem{Fayet:1978ig}
P.~Fayet, ``{Spontaneous Generation of Massive Multiplets and Central Charges
  in Extended Supersymmetric Theories},''
  \href{http://dx.doi.org/10.1016/0550-3213(79)90162-7}{{\em Nucl.Phys.} {\bf
  B149} (1979)  137}.

\bibitem{Cheung:2009dc}
C.~Cheung and D.~O'Connell, ``{Amplitudes and Spinor-Helicity in Six
  Dimensions},'' \href{http://dx.doi.org/10.1088/1126-6708/2009/07/075}{{\em
  JHEP} {\bf 0907} (2009)  075}, \href{http://arxiv.org/abs/0902.0981}{{\tt
  arXiv:0902.0981 [hep-th]}}.

\bibitem{CaronHuot:2010rj}
S.~Caron-Huot and D.~O'Connell, ``{Spinor Helicity and Dual Conformal Symmetry
  in Ten Dimensions},'' \href{http://arxiv.org/abs/1010.5487}{{\tt
  arXiv:1010.5487 [hep-th]}}.

\bibitem{Hatsuda:2008pm}
M.~Hatsuda, Y.-t. Huang, and W.~Siegel, ``{First-quantized N=4 Yang-Mills},''
  \href{http://dx.doi.org/10.1088/1126-6708/2009/04/058}{{\em JHEP} {\bf 0904}
  (2009)  058}, \href{http://arxiv.org/abs/0812.4569}{{\tt arXiv:0812.4569
  [hep-th]}}.

\bibitem{Elvang:2011fx}
H.~Elvang, Y.-t. Huang, and C.~Peng, ``{On-shell superamplitudes in $N<4$
  SYM},'' \href{http://arxiv.org/abs/1102.4843}{{\tt arXiv:1102.4843
  [hep-th]}}.

\bibitem{Cachazo:2004by}
F.~Cachazo, P.~Svrcek, and E.~Witten, ``{Gauge theory amplitudes in twistor
  space and holomorphic anomaly},''
  \href{http://dx.doi.org/10.1088/1126-6708/2004/10/077}{{\em JHEP} {\bf 10}
  (2004)  077},
\href{http://arxiv.org/abs/hep-th/0409245}{{\tt arXiv:hep-th/0409245}}.
%%CITATION = HEP-TH/0409245;%%.

\bibitem{Alday:2007hr}
L.~F. Alday and J.~M. Maldacena, ``{Gluon scattering amplitudes at strong
  coupling},'' \href{http://dx.doi.org/10.1088/1126-6708/2007/06/064}{{\em
  JHEP} {\bf 0706} (2007)  064}, \href{http://arxiv.org/abs/0705.0303}{{\tt
  arXiv:0705.0303 [hep-th]}}.

\bibitem{Bardeen:1995gk}
W.~A. Bardeen, ``{Selfdual Yang-Mills theory, integrability and multiparton
  amplitudes},''
{\em Prog. Theor. Phys. Suppl.} {\bf 123} (1996)  1--8.
%%CITATION = PTPSA,123,1;%%.

\bibitem{Selivanov:1996gw}
K.~G. Selivanov, ``{Multigluon Tree Amplitudes and Self-Duality Equation},''
\href{http://arxiv.org/abs/hep-ph/9604206}{{\tt arXiv:hep-ph/9604206}}.
%%CITATION = HEP-PH/9604206;%%.

\bibitem{Korepin:1996mm}
V.~E. Korepin and T.~Oota, ``{Scattering of plane waves in self-dual Yang-Mills
  theory},'' \href{http://dx.doi.org/10.1088/0305-4470/29/24/003}{{\em J.
  Phys.} {\bf A29} (1996)  L625--L628},
\href{http://arxiv.org/abs/hep-th/9608064}{{\tt arXiv:hep-th/9608064}}.
%%CITATION = HEP-TH/9608064;%%.

\bibitem{Adler:1964um}
S.~L. Adler, ``{Consistency conditions on the strong interactions implied by a
  partially conserved axial vector current},''
  \href{http://dx.doi.org/10.1103/PhysRev.137.B1022}{{\em Phys.Rev.} {\bf 137}
  (1965)  B1022--B1033}.

\bibitem{Bianchi:2008pu}
M.~Bianchi, H.~Elvang, and D.~Z. Freedman, ``{Generating Tree Amplitudes in
  $N=4$ SYM and N = 8 SG},''
  \href{http://dx.doi.org/10.1088/1126-6708/2008/09/063}{{\em JHEP} {\bf 09}
  (2008)  063},
\href{http://arxiv.org/abs/0805.0757}{{\tt arXiv:0805.0757 [hep-th]}}.
%%CITATION = 0805.0757;%%.

\bibitem{ArkaniHamed:2008gz}
N.~Arkani-Hamed, F.~Cachazo, and J.~Kaplan, ``{What is the Simplest Quantum
  Field Theory?},'' \href{http://dx.doi.org/10.1007/JHEP09(2010)016}{{\em JHEP}
  {\bf 09} (2010)  016},
\href{http://arxiv.org/abs/0808.1446}{{\tt arXiv:0808.1446 [hep-th]}}.
%%CITATION = 0808.1446;%%.

\bibitem{Kallosh:2008rr}
R.~Kallosh and T.~Kugo, ``{The footprint of E7 in amplitudes of N=8
  supergravity},'' \href{http://dx.doi.org/10.1088/1126-6708/2009/01/072}{{\em
  JHEP} {\bf 01} (2009)  072},
\href{http://arxiv.org/abs/0811.3414}{{\tt arXiv:0811.3414 [hep-th]}}.
%%CITATION = 0811.3414;%%.

\bibitem{Brodel:2009hu}
J.~Broedel and L.~J. Dixon, ``{$R^4$ counterterm and E7(7) symmetry in maximal
  supergravity},'' \href{http://dx.doi.org/10.1007/JHEP05(2010)003}{{\em JHEP}
  {\bf 05} (2010)  003},
\href{http://arxiv.org/abs/0911.5704}{{\tt arXiv:0911.5704 [hep-th]}}.
%%CITATION = 0911.5704;%%.

\bibitem{Elvang:2010kc}
H.~Elvang and M.~Kiermaier, ``{Stringy KLT relations, global symmetries, and
  $E_7(7)$ violation},'' \href{http://dx.doi.org/10.1007/JHEP10(2010)108}{{\em
  JHEP} {\bf 10} (2010)  108},
\href{http://arxiv.org/abs/1007.4813}{{\tt arXiv:1007.4813 [hep-th]}}.
%%CITATION = 1007.4813;%%.

\bibitem{Bossard:2010dq}
G.~Bossard, C.~Hillmann, and H.~Nicolai, ``{E7(7) symmetry in perturbatively
  quantised N=8 supergravity},''
  \href{http://dx.doi.org/10.1007/JHEP12(2010)052}{{\em JHEP} {\bf 1012} (2010)
   052}, \href{http://arxiv.org/abs/1007.5472}{{\tt arXiv:1007.5472 [hep-th]}}.

\bibitem{Bossard:2010bd}
G.~Bossard, P.~Howe, and K.~Stelle, ``{On duality symmetries of supergravity
  invariants},'' \href{http://dx.doi.org/10.1007/JHEP01(2011)020}{{\em JHEP}
  {\bf 1101} (2011)  020}, \href{http://arxiv.org/abs/1009.0743}{{\tt
  arXiv:1009.0743 [hep-th]}}.

\bibitem{Beisert:2010jx}
N.~Beisert {\em et al.}, ``{E7(7) constraints on counterterms in N=8
  supergravity},'' \href{http://dx.doi.org/10.1016/j.physletb.2010.09.069}{{\em
  Phys. Lett.} {\bf B694} (2010)  265--271},
\href{http://arxiv.org/abs/1009.1643}{{\tt arXiv:1009.1643 [hep-th]}}.
%%CITATION = 1009.1643;%%.

\bibitem{Elvang:2009wd}
H.~Elvang, D.~Z. Freedman, and M.~Kiermaier, ``{Solution to the Ward Identities
  for Superamplitudes},'' \href{http://dx.doi.org/10.1007/JHEP10(2010)103}{{\em
  JHEP} {\bf 1010} (2010)  103}, \href{http://arxiv.org/abs/0911.3169}{{\tt
  arXiv:0911.3169 [hep-th]}}.

\bibitem{Elvang:2008vz}
H.~Elvang, D.~Z. Freedman, and M.~Kiermaier, ``{Proof of the MHV vertex
  expansion for all tree amplitudes in $N=4$ SYM theory},''
  \href{http://dx.doi.org/10.1088/1126-6708/2009/06/068}{{\em JHEP} {\bf 06}
  (2009)  068},
\href{http://arxiv.org/abs/0811.3624}{{\tt arXiv:0811.3624 [hep-th]}}.
%%CITATION = 0811.3624;%%.

\bibitem{Kiermaier:2009yu}
M.~Kiermaier and S.~G. Naculich, ``{A Super MHV vertex expansion for N=4 SYM
  theory},'' \href{http://dx.doi.org/10.1088/1126-6708/2009/05/072}{{\em JHEP}
  {\bf 0905} (2009)  072}, \href{http://arxiv.org/abs/0903.0377}{{\tt
  arXiv:0903.0377 [hep-th]}}.

\bibitem{Bullimore:2010dz}
M.~Bullimore, ``{MHV Diagrams from an All-Line Recursion Relation},''
  \href{http://arxiv.org/abs/1010.5921}{{\tt arXiv:1010.5921 [hep-th]}}.

\bibitem{Risager:2005vk}
K.~Risager, ``{A Direct proof of the CSW rules},''
  \href{http://dx.doi.org/10.1088/1126-6708/2005/12/003}{{\em JHEP} {\bf 0512}
  (2005)  003}, \href{http://arxiv.org/abs/hep-th/0508206}{{\tt
  arXiv:hep-th/0508206 [hep-th]}}.

\bibitem{Elvang:2008na}
H.~Elvang, D.~Z. Freedman, and M.~Kiermaier, ``{Recursion Relations, Generating
  Functions, and Unitarity Sums in N=4 SYM Theory},''
  \href{http://dx.doi.org/10.1088/1126-6708/2009/04/009}{{\em JHEP} {\bf 0904}
  (2009)  009}, \href{http://arxiv.org/abs/0808.1720}{{\tt arXiv:0808.1720
  [hep-th]}}.

\bibitem{CaronHuot:2010ek}
S.~Caron-Huot, ``{Notes on the scattering amplitude / Wilson loop duality},''
  \href{http://arxiv.org/abs/1010.1167}{{\tt arXiv:1010.1167 [hep-th]}}.

\bibitem{Dixon:2010ik}
L.~J. Dixon, J.~M. Henn, J.~Plefka, and T.~Schuster, ``{All tree-level
  amplitudes in massless QCD},''
  \href{http://dx.doi.org/10.1007/JHEP01(2011)035}{{\em JHEP} {\bf 1101} (2011)
   035}, \href{http://arxiv.org/abs/1010.3991}{{\tt arXiv:1010.3991 [hep-ph]}}.

\bibitem{BoelsSchwinn}
R.~Boels and C.~Schwinn, ``On-shell supersymmetry for massive multiplets,''
  {\em to appear}  .

\end{thebibliography}
\providecommand{\href}[2]{#2}\begingroup\raggedright\endgroup

\end{document}